\documentclass[12pt]{article}
\usepackage[totalwidth=460truept,totalheight=624truept]{geometry}
\usepackage{latexsym,graphicx,amsfonts,amssymb,amsmath,xcolor,changepage,bm}
\usepackage[hypertexnames=false,hidelinks]{hyperref}

\def\theequation{\arabic{section}.\arabic{equation}}

\renewcommand{\theequation}{\thesection.\arabic{equation}}
\global\arraycolsep=1truept
\renewcommand{\theequation}{\arabic{section}.\arabic{equation}}
\allowdisplaybreaks
\null

\begin{document}

\bigskip

\begin{center}
{\Huge \textbf{\negthinspace Cosmological Inhomogeneities,}}

\vskip.6truecm

{\Huge \textbf{Primordial\ Black Holes,}}

\vskip.6truecm

{\Huge \textbf{and\ a Hypothesis}}

\vskip.65truecm

{\Huge \textbf{On\ the Death of the Universe}}

\vskip 1truecm

\textsl{Damiano Anselmi}

\vskip.1truecm

{\small \textit{Dipartimento di Fisica \textquotedblleft
E.Fermi\textquotedblright , Universit\`{a} di Pisa, Largo B.Pontecorvo 3,
56127 Pisa, Italy}}

{\small \textit{INFN, Sezione di Pisa, Largo B. Pontecorvo 3, 56127 Pisa,
Italy}}

{\small damiano.anselmi@unipi.it}

\vskip 1.5truecm

\textbf{Abstract}
\end{center}

We study the impact of the expansion of the universe on a broad class of
objects, including black holes, neutron stars, white dwarfs, and others.
Using metrics that incorporate primordial inhomogeneities, the effects of a
hypothetical \textquotedblleft center of the universe\textquotedblright\ on
inflation are calculated. Dynamic\ coordinates for black holes that account
for expansions or contractions with arbitrary rates are provided. We
consider the possibility that the universe may be bound to evolve into an
ultimate state of \textquotedblleft total dilution\textquotedblright ,
wherein stable particles are so widely separated that physical communication
among them will be impossible for eternity. This is also a scenario of
\textquotedblleft cosmic virtuality\textquotedblright , as no wave-function
collapse would occur again. We provide classical models evolving this way,
based on the Majumdar-Papapetrou geometries. More realistic configurations,
instead, indicate that gravitational forces locally counteract expansion,
except in the universe's early stages. We comment on whether quantum
phenomena may dictate that total dilution is indeed the cosmos' ultimate
destiny.

\vfill\eject

\section{Introduction}

\label{intro}\setcounter{equation}{0}

In a scenario where the expansion of the universe is accelerating and the
event horizon is located at a finite comoving distance, i.e., 
\begin{equation}
d(t)=\int_{t}^{\infty }\frac{\mathrm{d}t^{\prime }}{a(t^{\prime })}<\infty ,
\label{dcom}
\end{equation}%
where $a(t)$ is the scale factor, an admissible state is the one where all
the unstable particles have decayed, and the stable ones are so distant from
one another that they will be unable to exchange physical signals for
eternity. We label such a state \textquotedblleft total
dilution\textquotedblright . In a hypothetical situation of this type, the
particles are not actually \textquotedblleft particles\textquotedblright ,
but just wave functions that do not have the chance to be brought to reality
again by means of wave-function collapses, since they can no longer interact
with macroscopic bodies, like a detector, let alone meet an
\textquotedblleft observer\textquotedblright . Thus, the state of total
dilution is also a state of \textquotedblleft cosmic
virtuality\textquotedblright .

Whether the expansion of the universe will satisfy (\ref{dcom}) forever or
not is a topic of active ongoing research. Here we assume that it will, and
try to figure out if, in the long run, total dilution is the final state for
a generic set of initial conditions, i.e., there exists a time $t_{\text{dis}%
}$, such that, for every $t>t_{\text{dis}}$, every pair of particles is
separated by a comoving distance larger than $d(t)$. Is the expansion
ultimately going to expand everything, including the planetary systems, the
celestial bodies, maybe even the atoms?

The common lore is that the expansion has a negligible impact on scales
smaller than galaxy clusters, such as within planetary systems, where the
attractive force of gravity prevails. Yet, there are various configurations
where the gravitational force is compensated by opposing forces, and the
effects of the expansion are the only surviving ones. The simplest example
is a homogeneous distribution of matter at large scales. There, the
gravitational attraction exerted by a cluster A on a cluster B\ is
compensated by the force exerted on B by an opposite cluster C. If we add
isotropy to the list of assumptions, we end up with an FLRW metric, where
objects preserve their positions in comoving coordinates: the clusters drift
far away from one another, till they are unable to physically communicate.

One may think that homogeneity is crucial to have this result, but this is
not true: the gravitational attraction can be compensated by forces of a
different nature. A remarkable example is provided by the
Majumdar-Papapetrou system \cite{papapetrou}, which contemplates an
arbitrary distribution of extremal charged black holes, arranged so that the
gravitational attraction is balanced exactly by the electrostatic repulsion.
The metric can be generalized to incorporate a nonvanishing cosmological
constant $\Lambda $, as shown by Kastor and Traschen in \cite{KastorTraschen}%
. We obtain a nonhomogeneous system where the relative positions of the
black holes remain fixed in dynamic (comoving)\ coordinates: the black holes
fall apart indefinitely.

This raises the question whether the expansion prevails any time the
gravitational force is compensated by an opposing force. We study systems
like neutron stars, white dwarfs and black dwarfs, where gravity is balanced
by the fermion degeneracy pressure \cite{Stars,Landau5}. We demonstrate that
these celestial bodies do not expand with respect to the event horizon. They
would have expanded in the early stages of the universe, but no stars
existed at that time.

Precisely, the expansion of the universe adds a \textquotedblleft
centrifugal force\textquotedblright\ to the balance of forces. Once this
contribution is taken into account, one finds that the equations admit a
configuration of hydrodynamic equilibrium (which allows for the presence of
white dwarfs and neutron stars), only if the centrifugal force due to the
expansion is smaller than the gravitational force at the border of the star.
This condition has been fulfilled since the time of last scattering (and
sometime before), but not during inflation.

These results agree with the findings of Price and Romano \cite{Price} (see
also \cite{Faraoni}), who showed that simple weakly coupled systems, like a
classical \textquotedblleft atom\textquotedblright , either completely
follow the cosmological expansion, or completely ignore it.

So, although an expansion satisfying (\ref{dcom}) makes total dilution an
admissible state of the universe, it is not enough to reach it classically.
Quantum effects must be advocated for that to happen. Generically speaking,
an argument in favor is that the theory of quanta is incompatible with the
idea of absolute stability. Ways of escaping the gravitational attraction in
the extremely long run are the black-hole evaporation \cite{evaporation},
and possibly gravitational analogues \cite{gravSHES,Wilczek,Wondrak} of the
(Sauter-Heisenberg-Euler-)Schwinger pair production mechanism \cite{SHES},
which could even make \textquotedblleft everything
evaporate\textquotedblright . Still, the final verdict on whether total
dilution, or cosmic virtuality, is \textquotedblleft the fate of every
universe\textquotedblright\ awaits further investigations. We hope that the
results of this paper help assess the issue.

An accelerated expansion can be described by a positive cosmological
constant $\Lambda $, or, more generally, an inflaton field $\phi $ \cite%
{cosmology}. We study various systems of these types. For example, we
explore the possibility that the universe may have a \textquotedblleft
center\textquotedblright , or multiple centers, where perennial black holes
are concentrated. By incorporating such inhomogeneities into the history of
the early universe, we show that their impact on the power spectra of
primordial fluctuations leaves room for black holes of masses up to $10^{43}$%
kg, when the total mass of the observable universe is about $10^{53}$kg. We
work in the context of the Starobinsky scenario \cite{staro}, but the
arguments can be generalized to other types of inflation \cite{inflation}.

It is worth stressing that the primordial black holes we are talking about
are different from the ones commonly discussed in the literature \cite%
{primordialBH}: ours may have been there before inflation as well as through
it; instead, the ones considered in standard scenarios are supposed to form
in epochs that are posterior to inflation.

We also provide dynamic coordinates for static metrics. Among those, a
family of coordinates for Schwartschild and Schwartschild-de Sitter black
holes that account for expansions and contractions with arbitrary rates.

The paper is organized as follows. In section \ref{blackhole} we begin with
the last topic mentioned above, the study of dynamic coordinates for non
rotating black holes. In section \ref{papapetrou} we consider systems of the
Majumdar-Papapetrou and Kastor-Traschen types and show that they provide
exact toy models for the evolution toward total dilution. In section \ref%
{pbh} we study Starobinsky inflation with inhomogeneities, which may consist
of one or many primordial black holes. In section \ref{neutron} we discuss
the fate of neutron stars, white dwarfs and similar objects. In section \ref%
{fate} we discuss the fate of the universe in a broader sense. Section \ref%
{conclusions} contains the conclusions. Appendix \ref{higher} is devoted to
higher-order corrections to the black-hole metric of section \ref{pbh}.
Appendix \ref{fluidGR} contains a derivation of the equations of fluid
dynamics in general relativity. In appendix \ref{switch} we explain how to
switch from a set of particles to a fluid. In appendix \ref{dust} dust is
discussed in detail. Finally, appendix \ref{degeneracy}\ contains
derivations of the pressure and equations of state of an ideal degenerate
Fermi fluid \cite{Landau5}.

\section{Dynamic coordinates for black holes}

\label{blackhole}\setcounter{equation}{0}

In this section we provide dynamic coordinates for black holes, accounting
for expansions or contractions with arbitrary rates\footnote{%
For a review of the most popular coordinate choices for black holes, see 
\cite{Unruh}.}. We study the main physical properties and comment on their
significance for the investigation of this paper.

Consider\ the Schwartzschild metric%
\begin{equation}
\mathrm{d}s^{2}=g_{\mu \nu }\mathrm{d}x^{\mu }\mathrm{d}x^{\nu }=\left( 1-%
\frac{r_{g}}{r}\right) \mathrm{d}t^{2}-\frac{\mathrm{d}r^{2}}{1-\frac{r_{g}}{%
r}}-r^{2}(\mathrm{d}\theta ^{2}+\sin ^{2}\theta \hspace{0.01in}\hspace{0.01in%
}\mathrm{d}\varphi ^{2}),  \label{Schw}
\end{equation}%
the Schwarzschild-de Sitter metric 
\begin{equation}
\mathrm{d}s^{2}=\left( 1-\frac{r_{g}}{r}-H^{2}r^{2}\right) \mathrm{d}t^{2}-%
\frac{\mathrm{d}r^{2}}{1-\frac{r_{g}}{r}-H^{2}r^{2}}-r^{2}(\mathrm{d}\theta
^{2}+\sin ^{2}\theta \hspace{0.01in}\mathrm{d}\varphi ^{2})  \label{SchwC}
\end{equation}%
and the FLRW metric%
\begin{equation}
\mathrm{d}s^{2}=\mathrm{d}t^{2}-a(t)^{2}(\mathrm{d}r^{2}+r^{2}\mathrm{d}%
\theta ^{2}+r^{2}\sin ^{2}\theta \hspace{0.01in}\hspace{0.01in}\mathrm{d}%
\varphi ^{2})  \label{Fr}
\end{equation}%
at zero spatial curvature, where%
\begin{equation}
a(t)=a_{0}\mathrm{e}^{Ht}  \label{at}
\end{equation}%
is the scale factor, $r_{g}$ is the Schwartzschild radius, $H$ is the Hubble
parameter, assumed to be constant, and $a_{0}$ is another constant. We
search for a solution that

1) gives (\ref{Fr}) in the limit $r_{g}\rightarrow 0$;

2) gives (\ref{Schw}) for $r>r_{g}$ in the limit $H\rightarrow 0$;

3) satisfies the Einstein equations%
\begin{equation}
R_{\mu \nu }-\frac{1}{2}g_{\mu \nu }R-\Lambda g_{\mu \nu }=0  \label{Eeq}
\end{equation}%
with a cosmological constant $\Lambda =3H^{2}$. The metric (\ref{SchwC})
satisfies 2) and 3), but not 1), while (\ref{Fr}) satisfies 1) and 3), but
not 2).

If we introduce a scale factor $a_{\lambda }(t)$, such that%
\begin{equation*}
\dot{a}_{\lambda }(t)=\lambda a_{\lambda }(t),
\end{equation*}%
where $\lambda $ is an arbitrary constant, we can actually treat a larger
class of metrics at once. The factor $a_{\lambda }(t)$ can describe a
physical expansion (when $\lambda =H$), or an artificial choice of expanding
or contracting coordinates.

Inserting the ansatz%
\begin{equation}
\mathrm{d}s^{2}=f(u)\mathrm{d}t^{2}-\frac{a_{\lambda }(t)^{2}\mathrm{d}r^{2}%
}{f(u)}-a_{\lambda }(t)^{2}r^{2}(\mathrm{d}\theta ^{2}+\sin ^{2}\theta 
\hspace{0.01in}\hspace{0.01in}\mathrm{d}\varphi ^{2})  \label{ansatz}
\end{equation}
into (\ref{Eeq}), having defined $u=u(t,r)\equiv a_{\lambda }(t)r$, we
obtain the differential equation%
\begin{equation*}
\frac{uf^{\prime }}{f}(f^{2}+\lambda ^{2}u^{2})=f(1-f)+3u^{2}(\lambda
^{2}-H^{2}f)
\end{equation*}%
for $f(u)$. Its solutions are $f(u)=f_{\pm }(u)$, where 
\begin{equation}
f_{\pm }(u)=w(u)\pm \sqrt{w(u)^{2}+\lambda ^{2}u^{2}},\qquad w(u)=\frac{1}{2}%
\left( 1-\frac{r_{g}}{u}-H^{2}u^{2}\right) .  \label{fu}
\end{equation}

The results,%
\begin{equation}
\mathrm{d}s_{\pm }^{2}=f_{\pm }(u)\mathrm{d}t^{2}-\frac{a_{\lambda }(t)^{2}}{%
f_{\pm }(u)}\mathrm{d}r^{2}-u^{2}(\mathrm{d}\theta ^{2}+\sin ^{2}\theta 
\hspace{0.01in}\hspace{0.01in}\mathrm{d}\varphi ^{2}),  \label{dspm}
\end{equation}%
are related to the Schwarzschild-de Sitter metric (\ref{SchwC}) by the
changes of coordinates%
\begin{equation*}
t^{\prime }=t+\frac{\lambda }{2}\int^{r^{\prime }}\frac{u\hspace{0.01in}%
\mathrm{d}u}{w(u)f_{\pm }(u)},\qquad r^{\prime }=ra_{\lambda }(t),\qquad
\theta ^{\prime }=\theta ,\qquad \varphi ^{\prime }=\varphi ,
\end{equation*}%
where the primes refer to (\ref{SchwC}).

When $\lambda =H$ the worldline element $\mathrm{d}s_{+}$ satisfies the
points 1-3) listed above: the function $f_{+}(u)$ is identically one for $%
r_{g}=0$, which gives the FLRW metric (\ref{Fr}); moreover, $\mathrm{d}%
s_{+}^{2}$ returns the black-hole metric (\ref{Schw}) for $r>r_{g}$ when $%
H\rightarrow 0$.

The metrics (\ref{dspm}) have some interesting properties, which we now list.

The horizons of the metric (\ref{SchwC}) are located at distances $r_{\ast }$
where $w(r_{\ast })=0$. This condition admits two solutions (event horizon
and cosmological horizon) for $Hr_{g}<2/3^{4/3}$ and no solution for $%
Hr_{g}>2/3^{4/3}$ (see fig. \ref{horizons}). The metrics $\mathrm{d}s_{\pm
}^{2}$ with $\lambda \neq 0$ do not have singularities away from $r\neq 0$.
In some sense, the parameter $\lambda $ acts as a \textquotedblleft
regulator\textquotedblright\ of the horizons. When $\lambda =H$, the
regulator is the cosmological constant. When $\lambda \neq H$, it is an
artificial choice of coordinates. The metrics $\mathrm{d}s_{+}^{2}$ with $%
H=0 $, $\lambda \neq 0$ can be used to reach the interior region of a
Schwartzschild black hole. See below for the amount of time a falling body
takes to cross the horizon.

Note that the function $f_{+}(u)$ is always positive, while the function $%
f_{-}(u)$ is always negative: $t$ and $r$ exchange their roles in the metric 
$\mathrm{d}s_{-}^{2}$.

When $\lambda $ tends to zero, the metric $\mathrm{d}s_{+}^{2}$ tends to (%
\ref{SchwC}) for $w(r)>0$, while the function $f_{+}(u)$ tends to zero for $%
w(r)<0$. The metric $\mathrm{d}s_{-}^{2}$ tends to (\ref{SchwC}) for $w(r)<0$%
, while $f_{-}(u)$ tends to zero for $w(r)>0$.

From now on, we focus on~$\mathrm{d}s_{+}^{2}$. For $w(u)>0$ we obtain the
expansion%
\begin{equation}
f_{+}(u)=2w(u)\left[ 1+\frac{u^{2}\lambda ^{2}}{4w(u)^{2}}-\frac{%
u^{4}\lambda ^{4}}{16w(u)^{4}}+\mathcal{O}\left( \lambda ^{6}\right) \right]
\label{fpexpa}
\end{equation}%
in powers of $\lambda $, which can be used away from the horizons. For $%
w(u)<0$ the expansion is $2w(u)$ minus the right hand side of (\ref{fpexpa}).

The function $f_{+}(u)$ tends to $\lambda ^{2}/H^{2}$ for $u\rightarrow
\infty $. Its expansion around $u=0$ reads%
\begin{equation*}
f(u)=\frac{\hat{u}u^{2}\lambda ^{2}}{1-\bar{u}}+\mathcal{O}(u^{6}),\qquad 
\hat{u}\equiv \frac{u}{r_{g}},
\end{equation*}%
which highlights the black-hole singularity.

\subsection{Light propagation}

Setting $\mathrm{d}s_{+}^{2}=0$ and $\mathrm{d}\theta =\mathrm{d}\varphi =0$%
, we can study the radial propagation of light. We obtain%
\begin{equation}
t=\int^{u}\frac{\mathrm{d}u^{\prime }}{\lambda u^{\prime }\pm
f_{+}(u^{\prime })}.  \label{light}
\end{equation}%
The plus sign in front of $f_{+}$ gives the trajectory of an emerging ray ($%
\mathrm{d}r/\mathrm{d}t>0$), while the minus sign gives the trajectory of a
ray moving toward the center ($\mathrm{d}r/\mathrm{d}t<0$).

The denominator of the integrand can potentially vanish on the
\textquotedblleft regularized horizons\textquotedblright , i.e., the values
of $u$ such that $w(u)=0$. Precisely, the amount of time spent around $%
w(u)\simeq 0$ is%
\begin{equation*}
\delta t\simeq \int^{u}\frac{\mathrm{d}u^{\prime }}{(\lambda \pm |\lambda
|)u^{\prime }\pm w(u^{\prime })+\mathcal{O}(w(u^{\prime })^{2})}.
\end{equation*}%
When $\lambda >0$, light takes an infinite amount of time to reach the
regularized horizons from the outside, while it escapes from the inside in a
finite amount of time. The opposite situations occur for $\lambda <0$.

\subsection{Motion of massive bodies}

Now we consider the trajectory of a body of mass $m$. With no loss of
generality, we can restrict to the plane $\theta =\pi /2$. We first describe
the motion in the $u$ variable. The Hamilton-Jacobi equation%
\begin{equation*}
g^{\mu \nu }(\partial _{\mu }S)(\partial _{\nu }S)=m^{2}
\end{equation*}%
for the action $S$ is solved by%
\begin{equation*}
S=-Et+L\varphi +F_{\pm }(u),
\end{equation*}%
where $E$ is the energy, $L$ is the angular momentum and the functions $%
F_{\pm }(u)$ have derivatives 
\begin{equation}
F_{\pm }^{\prime }(u)=\frac{1}{2w(u)}\left[ \pm \sqrt{E^{2}-2w(u)\left(
m^{2}+\frac{L^{2}}{u^{2}}\right) }-\frac{\lambda Eu}{f_{+}(u)}\right] .
\label{fpu}
\end{equation}%
The solutions of the equations of motion for the orbits are%
\begin{equation}
t=\frac{\partial F_{\pm }(u)}{\partial E},\qquad \varphi =-\frac{\partial
F_{\pm }(u)}{\partial L}.  \label{equa}
\end{equation}%
The limit $m\rightarrow 0$ returns the propagation of light. For example, at 
$L=m=0$ ($E>0$) the first equation of (\ref{equa}) gives back (\ref{light}).

\begin{figure}[t]
\begin{center}
\includegraphics[width=14truecm]{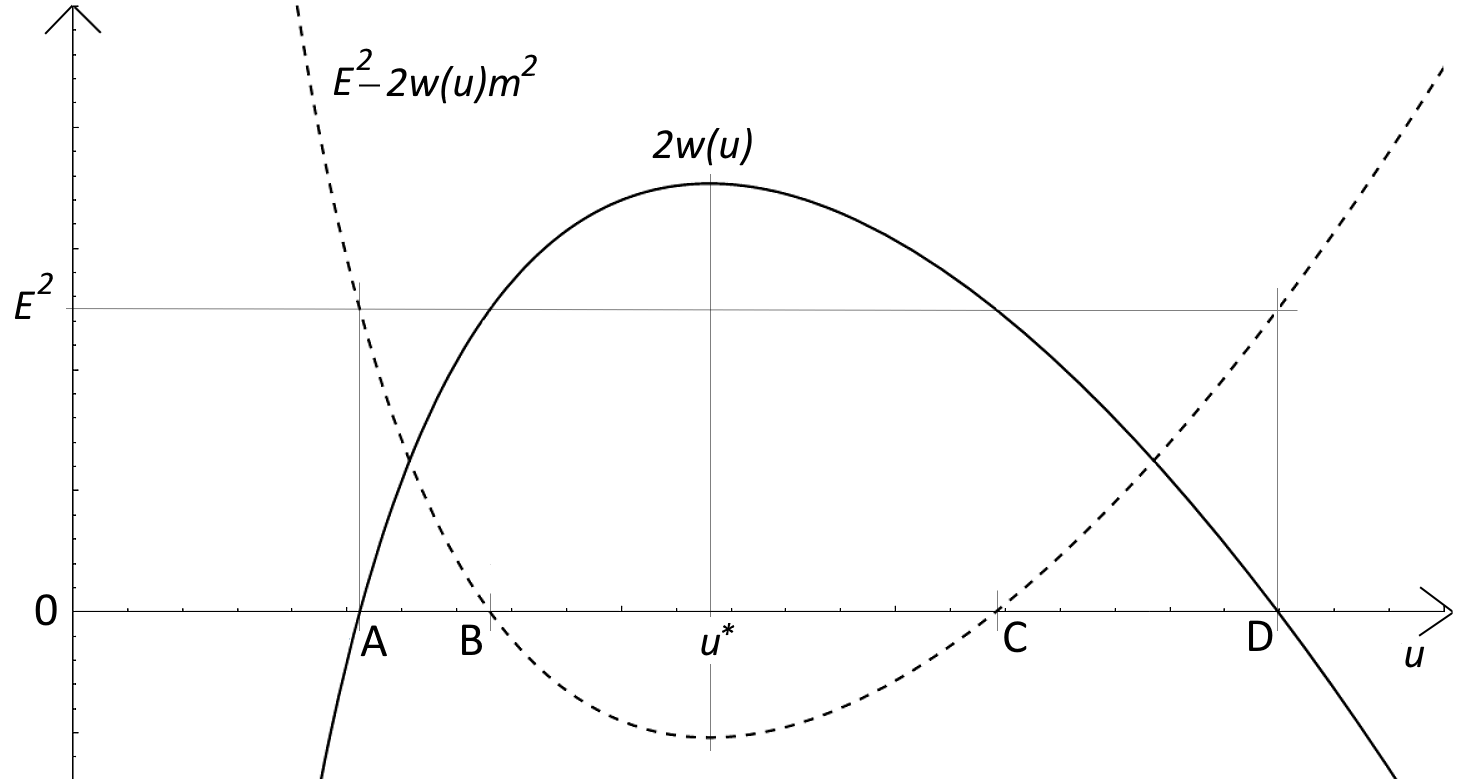}
\end{center}
\caption{Horizons and turning points for radially moving bodies}
\label{horizons}
\end{figure}

More generally, the following situations occur for $E>0$, $\lambda >0$:

1) $\mathrm{d}u/\mathrm{d}t$ is never singular and always positive away from 
$r\neq 0$ in the case $F_{+}$;

2) $\mathrm{d}u/\mathrm{d}t$ has the same sign as $-w(u)$, and is singular
only on the regularized horizons in the case $F_{-}$.

\noindent For $E>0$, $\lambda <0$ we have the following situations:

3) $\mathrm{d}u/\mathrm{d}t$ is never singular and always negative away from 
$r\neq 0$ in the case $F_{-}$;

4) $\mathrm{d}u/\mathrm{d}t$ has the same sign as $w(u)$ and is singular
only on the regularized horizons in the case $F_{+}$.

The potential singularities at $w(u)=0$ disappear in the cases 1) and 3),
because they are canceled by the expression within the square brackets of (%
\ref{fpu}).

When $E<0$ we have the same classification with $F_{+}\leftrightarrow F_{-}$.

Consider again the radial motion ($L=0$). In fig. \ref{horizons} we plot the
function $2w(u)$, which identifies the regularized horizons A and D, and the
argument $E^{2}-2m^{2}w(u)$ of the square root of (\ref{fpu}), which
identifies the turning points B and C of the trajectory, if any. We have
chosen typical values of $r_{g}$, $H$ and $m$. Higher energies translate the
second curve upward and eventually eliminate the turning points.

Assuming $\lambda >0$, consider the situation depicted in the figure. A
massive body emerging from the center $r=0$ can be described by the solution
with $F_{-}$ or the one with $F_{+}$. In the first case, it reaches the
regularized horizon A in the infinite future. In the second case, it crosses
A in a finite amount of time. Then, it reaches the turning point B, where $%
\mathrm{d}u/\mathrm{d}t$ vanishes, also in a finite amount of time. After
that, its trajectory is no longer described by the solution with $F_{+}$: we
need to switch to the solution with $F_{-}$, where we see the body falling
from B toward the regularized horizon A, which it reaches in the infinite
future.

A body coming from the regularized horizon D in the infinite past can either
move away to infinity in the infinite future (solution with $F_{-}$) or
first fall toward C (again solution with $F_{-}$), then turn, emerge from C,
cross D in a finite amount of time and proceed to infinity (solution with $%
F_{+}$).

If $E$ is large enough, there are no turning points B and C. Then an
emerging body (solution with $F_{+}$) crosses the regularized horizons in
finite amounts of time and proceeds to infinity. The solutions with $F_{-}$
are: body from D in the infinite past to A in the infinite future; body
emerging from the origin and reaching A in the infinite future; body
emerging from D in the infinite past and proceeding to infinity.

We can switch to the description of the motion in the variable $r$ by noting
that the velocity $\mathrm{d}r/\mathrm{d}t$ is related to $\mathrm{d}u/%
\mathrm{d}t$ by the formula%
\begin{equation*}
a_{\lambda }\frac{\mathrm{d}r}{\mathrm{d}t}=\frac{\mathrm{d}u}{\mathrm{d}t}%
-\lambda u.
\end{equation*}%
The solution with $F_{+}$,%
\begin{equation*}
a_{\lambda }\frac{\mathrm{d}r}{\mathrm{d}t}=\left( \frac{\partial
F_{+}^{\prime }(u)}{\partial E}\right) ^{-1}-\lambda u,
\end{equation*}%
has an inversion point $\bar{u}$ (where $\mathrm{d}r/\mathrm{d}t=0$) if the
condition $E=m\sqrt{f_{+}(\bar{u})}$ admits solutions.

Although the radial variable is $r$, the main properties of the solutions
involve $u$. For example, the regularized horizons A and D, as well as the
turning points B and C, as well as the inversion points $\mathrm{d}r/\mathrm{%
d}t=0$, occur for values of $u$ that are solely determined by the constants $%
r_{g}$, $H$, $E$ and $m$. It is also easy to derive circular orbits with $u=$
constant (see subsection \ref{staticco}). Since $u=a_{\lambda }r$, when $%
a_{\lambda }$ grows $r$ decreases correspondingly. The story does not change
when the system is quantized. If we study the energy levels of a particle in
the black-hole field, we find that the average distance from the center is $%
u $ = constant. In particular, the physical case $\lambda =H$ shows that a
particle interacting with an attractor remains bound to it indefinitely. The
expansion of the universe does not make a significant difference.

\subsection{Dynamic coordinates for charged black holes}

What said extends straightforwardly to charged black holes, where it is
sufficient to take%
\begin{equation}
w(u)=\frac{1}{2}\left( 1-\frac{r_{g}}{u}+G\frac{q^{2}}{u^{2}}%
-H^{2}u^{2}\right) ,  \label{chargedw}
\end{equation}%
$G$ being Newton's constant. With this $w(u)$, the solutions of the
Einstein-Maxwell equations%
\begin{equation}
R_{\mu \nu }-\frac{1}{2}g_{\mu \nu }R-\Lambda g_{\mu \nu }=8\pi GT_{\mu \nu
},\qquad D^{\mu }F_{\mu \nu }=0,  \label{EMeq}
\end{equation}%
in dynamic coordinates are given by the metrics (\ref{dspm}) together with
the vector potential 
\begin{equation*}
A_{\mu }=\left( \frac{q}{\sqrt{4\pi }u},0,0,0\right) ,
\end{equation*}%
where 
\begin{equation*}
T_{\mu \nu }=-F_{\mu \rho }F_{\nu }^{\hspace{0.02in}\hspace{0.03in}\rho }+%
\frac{g_{\mu \nu }}{4}F_{\alpha \beta }F^{\alpha \beta },\qquad F_{\mu \nu
}=\partial _{\mu }A_{\nu }-\partial _{\nu }A_{\mu },
\end{equation*}%
are the energy-momentum tensor and the field strength, respectively.

It is more challenging to arrange meaningful dynamic coordinates for
rotating black holes \cite{kerr}, since the rotation interferes with the
dynamics of the coordinate system.

\subsection{Dynamic isotropic coordinates}

\label{isoptropic}

In isotropic coordinates, the line element of the black hole without a
cosmological constant is \cite{Weyl} 
\begin{equation}
\mathrm{d}s^{2}=\frac{\left( 1-\frac{r_{g}}{4r}\right) ^{2}}{\left( 1+\frac{%
r_{g}}{4r}\right) ^{2}}\mathrm{d}t^{2}-\left( 1+\frac{r_{g}}{4r}\right) ^{4}(%
\mathrm{d}r^{2}+r^{2}\mathrm{d}\theta ^{2}+r^{2}\sin ^{2}\theta \hspace{%
0.01in}\mathrm{d}\varphi ^{2}).  \label{iso}
\end{equation}%
One can switch from (\ref{Schw}) to (\ref{iso}) by means of the change of
variables%
\begin{equation}
r^{\prime }=r\left( 1+\frac{r_{g}}{4r}\right) ^{2},  \label{rp}
\end{equation}%
where the primes refer to the Schwartzschild side. To the first order in $%
r_{g}$, the form (\ref{iso}) of the metric gives the common Newtonian
approximation%
\begin{equation}
\mathrm{d}s^{2}\simeq \left( 1-\frac{r_{g}}{r}\right) \mathrm{d}t^{2}-\left(
1+\frac{r_{g}}{r}\right) (\mathrm{d}x^{2}+\mathrm{d}y^{2}+\hspace{0.01in}%
\mathrm{d}z^{2})  \label{Newt}
\end{equation}%
at large distances.

It is surprisingly simple to promote (\ref{iso}) to a solution of the
Einstein equations (\ref{Eeq}) in the presence of a cosmological constant $%
\Lambda $: we just need to replace $r$ with $a(t)r$ and $\mathrm{d}r$ with $%
a(t)\mathrm{d}r$. This gives the McVittie metric \cite{McVittie}%
\begin{equation}
\mathrm{d}s^{2}=\frac{\left( 1-\frac{r_{g}}{4ra(t)}\right) ^{2}}{\left( 1+%
\frac{r_{g}}{4ra(t)}\right) ^{2}}\mathrm{d}t^{2}-\left( 1+\frac{r_{g}}{4ra(t)%
}\right) ^{4}a(t)^{2}(\mathrm{d}r^{2}+r^{2}\mathrm{d}\theta ^{2}+r^{2}\sin
^{2}\theta \hspace{0.01in}\mathrm{d}\varphi ^{2}),  \label{dsn}
\end{equation}%
where $a(t)$ is the one of (\ref{at}), with $H=\sqrt{\Lambda /3}$. The
coordinate change from (\ref{SchwC}) to (\ref{dsn}) is%
\begin{equation}
t^{\prime }=t+\int^{ra(t)}\frac{Hu\hspace{0.01in}\mathrm{d}u}{\Upsilon
(u)-H^{2}u^{2}},\qquad r^{\prime }=ra(t)\left( 1+\frac{r_{g}}{4ra(t)}\right)
^{2},\qquad \Upsilon (u)=\frac{\left( 1-\frac{r_{g}}{4u}\right) ^{2}}{\left(
1+\frac{r_{g}}{4u}\right) ^{6}},  \label{ch1}
\end{equation}%
where the primes refer to (\ref{SchwC}).

The static isotropic coordinates are%
\begin{equation}
\mathrm{d}s^{2}=\left( 1-\frac{r_{g}}{r^{\prime }(r)}-H^{2}r^{\prime \hspace{%
0.01in}2}(r)\right) \mathrm{d}t^{2}-\frac{r^{\prime \hspace{0.01in}2}(r)}{%
r^{2}}(\mathrm{d}r^{2}+r^{2}\mathrm{d}\theta ^{2}+r^{2}\sin ^{2}\theta 
\hspace{0.01in}\mathrm{d}\varphi ^{2}),  \label{statiso}
\end{equation}%
where the function $r^{\prime }(r)$ is the inverse of%
\begin{equation}
r(r^{\prime })=\exp \left( \int^{r^{\prime }}\frac{\hspace{0.01in}\mathrm{d}%
\tilde{r}}{\sqrt{\tilde{r}^{2}-r_{g}\tilde{r}-H^{2}\tilde{r}^{4}}}\right) .
\label{ch2}
\end{equation}%
The change of variables that relates (\ref{statiso}) to (\ref{dsn}) is the
composition of (\ref{ch1}) and (\ref{ch2}).

We see that isotropic coordinates are better suited for an easy switch from
the case of a vanishing cosmological constant to the one with a nonvanishing
cosmological constant. This is going to be useful in the next sections.

\subsection{Multiple black-hole systems}

We can generalize the Newton approximation (\ref{Newt}) by including an
arbitrary distribution of inhomogeneities, described by a potential $\Phi
(x,y,z)$ with vanishing Laplacian. In isotropic dynamic coordinates, the
result is 
\begin{equation}
\mathrm{d}s^{2}=\left( \!1-\frac{\Phi (x,y,z)}{a(t)}\right) \!\mathrm{d}%
t^{2}-\left( \!1+\frac{\Phi (x,y,z)}{a(t)}\right) \!a(t)^{2}(\mathrm{d}%
x^{2}\!+\!\mathrm{d}y^{2}\!+\!\hspace{0.01in}\mathrm{d}z^{2})+\mathcal{O}%
(\Phi ^{2}),  \label{dsnmulti}
\end{equation}%
to the first order in $\Phi $. For example, if we take%
\begin{equation}
\Phi (x,y,z)=\sum_{i}\frac{r_{ig}}{|\mathbf{r}-\mathbf{r}_{i}|},  \label{Phi}
\end{equation}%
where $\mathbf{r}=(x,y,z)$, $\mathbf{r}_{i}$ are the black-hole positions
and $r_{ig}$ are their Schwartzschild radii, formula (\ref{dsnmulti}) shows
that the singularities remain in the same positions $\mathbf{r}=\mathbf{r}%
_{i}$, so the expansion does not affect the relative comoving distances $|%
\mathbf{r}_{i}-\mathbf{r}_{j}|$. Instead, the expansion acts on the
potential $\Phi (x,y,z)$ as a whole, by means on the scaling factor $a(t)$
that divides $\Phi $.

This configuration aligns with our desired outcome, that is to say, a system
expanding along with the universe. However, it is not satisfactory, because
it merely assumes the result. The black holes do not stay fixed at $\mathbf{r%
}=\mathbf{r}_{i}$ by themselves: a force of unspecified nature must be
advocated to compensate for their gravitational attraction. In the following
section we present a system where the missing force is explicitly identified.

\section{Multi extremal black-hole systems}

\label{papapetrou}\setcounter{equation}{0}

The Majumdar-Papapetrou system contemplates an arbitrary distribution of
extremal charged black holes. The charges $q_{i}=$ $r_{ig}/(2\sqrt{G})$ are
adjusted to counterbalance the gravitational attractions by means of
electrostatic repulsions. Referring to formula (\ref{chargedw}), the
parenthesis becomes a perfect square in the case of a single black hole at $%
H=0$.

The system is described by the solution \cite{papapetrou}%
\begin{equation}
\mathrm{d}s^{2}=\left( 1+\frac{\Phi }{2}\right) ^{-2}\mathrm{d}t^{2}-\left(
1+\frac{\Phi }{2}\right) ^{2}(\mathrm{d}x^{2}+\mathrm{d}y^{2}+\hspace{0.01in}%
\mathrm{d}z^{2}),\qquad A_{\mu }=\frac{1}{\sqrt{4\pi G}}\frac{\Phi }{2+\Phi }%
(1,0,0,0).  \label{petro}
\end{equation}%
of the Einstein-Maxwell equations (\ref{EMeq}) without a cosmological
constant, where, as above, $\Phi =\Phi (x,y,z)$ is a function with vanishing
Laplacian\footnote{%
A rotating generalization is also known, given by the Perj\'{e}%
s-Israel-Wilson metrics \cite{PIW}.}.

Following the strategy outlined so far, it is easy to generalize this system
to a solution of (\ref{EMeq}) with a nonvanishing cosmological constant $%
\Lambda $. It is sufficient to rescale $\mathrm{d}x$, $\mathrm{d}y$ and $%
\hspace{0.01in}\mathrm{d}z$ by $a(t)$ and $\Phi $ by $a(t)^{-1}$, where the
scaling factor $a(t)$ is the one of (\ref{at}), with $H=\sqrt{\Lambda /3}$.
We obtain%
\begin{eqnarray}
\mathrm{d}s^{2} &=&\left( 1+\frac{\Phi }{2a(t)}\right) ^{-2}\mathrm{d}%
t^{2}-\left( 1+\frac{\Phi }{2a(t)}\right) ^{2}a(t)^{2}(\mathrm{d}x^{2}+%
\mathrm{d}y^{2}+\hspace{0.01in}\mathrm{d}z^{2}),  \notag \\
A_{\mu } &=&\frac{1}{\sqrt{4\pi G}}\frac{\Phi }{2a(t)}\left( 1+\frac{\Phi }{%
2a(t)}\right) ^{-1}(1,0,0,0).  \label{papadyn}
\end{eqnarray}%
This extension was already noted by Kastor and Traschen \cite{KastorTraschen}%
. To order one in $G$, the metric coincides with (\ref{dsnmulti}), while the
vector potential $A_{\mu }$ provides the compensating force advocated in the
previous section.

Choosing (\ref{Phi}), we see, once again, that the relative positions of the
black holes in dynamic coordinates do not change during the expansion of the
universe. More explicitly, consider the motion of an object of mass $m$ and
charge $q$ in the Maxwell/gravitational field (\ref{papadyn}). Instead of
solving the Hamilton-Jacobi equation, it is simpler to study the geodesics%
\begin{equation}
\frac{\mathrm{d}U^{\mu }}{\mathrm{d}s}+\Gamma _{\nu \rho }^{\mu }U^{\nu
}U^{\rho }=\frac{q}{m}\sqrt{4\pi }F_{\hspace{0.03in}\nu }^{\mu }U^{\nu },
\label{geodesic}
\end{equation}%
where $U^{\mu }=\mathrm{d}x^{\mu }(s)/\mathrm{d}s$ is the four-velocity, $%
x^{\mu }(s)$ denotes the trajectory of the body and $\mathrm{d}s=\sqrt{%
g_{\mu \nu }\mathrm{d}x^{\mu }\mathrm{d}x^{\nu }}$ is the worldline element.
It is easy to check that a solution of (\ref{geodesic}) is $U^{\mu
}=(1,0,0,0)/\sqrt{g_{00}}=(\mathrm{d}t/\mathrm{d}s,0,0,0)$, as long as the
moving body is extremal as well, i.e., $q=m\sqrt{G}$. Indeed, the equations (%
\ref{geodesic}) with $\mu =i$ give%
\begin{equation*}
-\frac{\partial _{i}g_{00}}{2g_{00}}=-\frac{q\partial _{i}\sqrt{g_{00}}}{m%
\sqrt{G}\sqrt{g_{00}}},
\end{equation*}%
which is true for $q=m\sqrt{G}$. Instead, the equation (\ref{geodesic}) with 
$\mu =0$ gives the identity%
\begin{equation*}
\frac{\mathrm{d}}{\mathrm{d}s}\frac{1}{\sqrt{g_{00}}}+\frac{\partial
_{0}g_{00}}{2g_{00}^{2}}=\frac{1}{\sqrt{g_{00}}}\partial _{0}\frac{1}{\sqrt{%
g_{00}}}+\frac{\partial _{0}g_{00}}{2g_{00}^{2}}=0.
\end{equation*}

We conclude that an extremal black hole at rest in a system of extremal
black holes (also at rest) does not feel the presence of the others. Then
the expansion prevails and the system (assuming that it is made of
hypothetical elementary particles) ultimately evolves into the state of
total dilution\footnote{%
Strictly speaking, the singularities of $\Phi $ at $\mathbf{r}=\mathbf{r}%
_{i} $ make the physical distance between each pair of centers infinite,
even before reaching the state of total dilution. The reason is that we are
considering idealized pointlike objects. We can circumvent the difficulty by
considering spheres of radii $r_{ig}$\ around the black holes, and defining
the state of total dilution as the one where such spheres can no longer
physically communicate with one another.}. This raises the question whether
the same fate awaits any system where the gravitational attraction is
balanced by an opposing force of different nature.

In fact, the extremality condition $q=m\sqrt{G}$ relating the mass $m$ and
the charge $q$ is not satisfied by the known elementary particles. For
example, the electron has $q/(m\sqrt{G})\simeq 10^{21}$. We can imagine
almost neutral celestial bodies, each having a slight excess of positively
charged particles and such that its total mass $M$ is indeed equal to its
total charge $Q$ multiplied by $\sqrt{G}$. A system of such bodies evolves
according to the dynamics described here\footnote{%
We are tacitly assuming that we can treat the bodies as pointlike.
Corrections due to their extensions are present, but they are not expected
to change the ultimate outcome of the dynamics.}.

To conclude, the solution (\ref{papadyn}) describes a toy model that
embodies the proposal we are advocating (the evolution toward the ultimate
state of total dilution). However, it is not realistic. In section \ref%
{neutron} we discuss how general its dynamics is, within the context of
realistic systems.

\section{Inflation with primordial inhomogeneities}

\label{pbh}\setcounter{equation}{0}

In this section we consider the expansion due to an inflaton field $\phi $.
For concreteness, we focus on the Starobinsky scenario, defined by the
potential%
\begin{equation}
V(\phi )=\frac{m_{\phi }^{2}}{2\hat{\kappa}^{2}}\left( 1-\mathrm{e}^{\hat{%
\kappa}\phi }\right) ^{2},  \label{staropot}
\end{equation}%
where $\hat{\kappa}=\sqrt{16\pi G/3}$ and $m_{\phi }$ is the inflaton mass.
Different potentials can be treated along the same lines. The action is%
\begin{equation}
S=-\frac{1}{16\pi G}\int \mathrm{d}^{4}x\sqrt{-g}R+\frac{1}{2}\int \mathrm{d}%
^{4}x\sqrt{-g}\left[ \left( \partial _{\mu }\phi )g^{\mu \nu }\partial _{\nu
}\phi -2V(\phi )\right) \right] ,  \label{S}
\end{equation}%
and the field equations read%
\begin{equation}
R_{\mu \nu }-\frac{1}{2}g_{\mu \nu }R=8\pi GT_{\mu \nu },\qquad \frac{1}{%
\sqrt{-g}}\partial _{\mu }(\sqrt{-g}g^{\mu \nu }\partial _{\nu }\phi
)+V^{\prime }(\phi )=0,  \label{staroeq}
\end{equation}%
where 
\begin{equation*}
T_{\mu \nu }=(\partial _{\mu }\phi )(\partial _{\nu }\phi )-\frac{1}{2}%
g_{\mu \nu }g^{\alpha \beta }(\partial _{\alpha }\phi )(\partial _{\beta
}\phi )+g_{\mu \nu }V(\phi )
\end{equation*}%
is the energy-momentum tensor.

We start from the homogeneous solution and then work out the corrections
that account for the presence of primordial inhomogeneities, in the form of
black holes or heavy massive bodies.

In the homogeneous case, the FLRW metric is (\ref{Fr}), but $a(t)$ is not
the one of (\ref{at}). Rather, the Hubble parameter $H\equiv \dot{a}/a$ is
time dependent, and so is the inflaton field $\phi =\phi _{0}(t)$. The
solution can be encoded into the \textquotedblleft running
coupling\textquotedblright\ 
\begin{equation}
\alpha =\sqrt{\frac{4\pi G}{3}}\frac{\dot{\phi}_{0}}{H}=\sqrt{-\frac{\dot{H}%
}{3H^{2}}},  \label{alfa}
\end{equation}%
of the \textquotedblleft cosmic RG flow\textquotedblright\ \cite{CMBrunning}%
, defined by the \textquotedblleft beta function\textquotedblright 
\begin{equation}
\beta _{\alpha }\equiv \frac{\mathrm{d}\alpha }{\mathrm{d\ln }|\tau |}%
=-2\alpha ^{2}f(\alpha ),  \label{beta}
\end{equation}%
where $\tau $ denotes the conformal time, 
\begin{equation}
\tau =-\int_{t}^{+\infty }\frac{\mathrm{d}t^{\prime }}{a(t^{\prime })}%
,\qquad \frac{\mathrm{d}}{\mathrm{d}t}=\frac{H}{aH\tau }\frac{\mathrm{d}}{%
\mathrm{d\ln }|\tau |},  \label{tau}
\end{equation}%
and $f(\alpha )$ is a known function of $\alpha $. Here it is sufficient to
expand $f(\alpha )$ in powers of $\alpha $, as it is for $H$ and the other
basic quantities:%
\begin{eqnarray}
f(\alpha ) &=&1+\frac{5}{6}\alpha +\frac{25}{9}\alpha ^{2}+\frac{383}{27}%
\alpha ^{3}+\frac{8155}{81}\alpha ^{4}+\mathcal{O}(\alpha ^{5}),  \notag \\
H &=&\frac{m_{\phi }}{2}\left( 1-\frac{3}{2}\alpha +\frac{7}{4}\alpha ^{2}-%
\frac{47}{24}\alpha ^{3}+\frac{293}{144}\alpha ^{4}\right) +\mathcal{O}%
(\alpha ^{5}),  \notag \\
-aH\tau &=&1+3\alpha ^{2}+12\alpha ^{3}+91\alpha ^{4}+\mathcal{O}(\alpha
^{5}).  \label{H}
\end{eqnarray}%
The expansion of $\dot{\phi}_{0}$ follows form (\ref{alfa}) and the one of $%
H $. The scaling factor $a(t)$ can be derived from $H$. The expansion of $%
\phi _{0}$ is rarely needed.

Comparing the predictions on the scalar spectra with observational data, one
finds $\alpha \simeq 1/115$, so in most situations it is enough to
concentrate on the first corrections listed above, or even set $\alpha
\simeq 0$.

The RG\ interpretation of inflation, where $\alpha $ is another way of
encoding the usual slow-roll parameter, has some advantages. For example,
the power spectra satisfy Callan-Symanzik equations in the superhorizon
limit. The other advantage is that we can treat the expansions in powers of $%
\alpha $ systematically (see for example \cite{CMBrunning2}), which is going
to be useful in a moment.

We want to show that we can include\ inhomogeneities (in the form of a
single black hole, for now) into an extended metric and an extended inflaton
field, still solving the equations (\ref{staroeq}). We search for the
solution in isotropic dynamic coordinates and work it out as an expansion in
powers of three quantities, that is to say,%
\begin{equation}
\alpha ,\qquad \frac{r_{g}}{u},\qquad \frac{1}{m_{\phi }^{2}u^{2}},
\label{param}
\end{equation}%
where $u=ra(t)$. As long as $u$ is larger than the Schwartzschild radius $%
r_{g}$, the parameters (\ref{param}) are small in the physical situations we
have in mind.

We organize the expansion in multiple tiers. The primary expansion is in
powers of $r_{g}/u$. Its coefficients undergo a second-tier expansion in
powers of $1/(m_{\phi }u)$, whose coefficients, in turn, undergo a
third-tier expansion in powers of $\alpha $.

The zeroth order in $r_{g}/u$ is just the metric (\ref{Fr}) with the scaling
factor $a(t)$ and the inflaton field $\phi _{0}(t)$ implied by (\ref{H}).
The first order in $r_{g}/u$ is independent of $1/(m_{\phi }^{2}u^{2})$: 
\begin{eqnarray}
\mathrm{d}s^{2} &=&\left( 1-\frac{2r_{g}H}{m_{\phi }u}\right) \mathrm{d}%
t^{2}-\left( 1+\frac{2r_{g}H}{m_{\phi }u}\right) a(t)^{2}(\mathrm{d}%
r^{2}+r^{2}\mathrm{d}\theta ^{2}+r^{2}\sin ^{2}\theta \hspace{0.01in}\mathrm{%
d}\varphi ^{2})+\mathcal{O}\left( \frac{r_{g}^{2}}{u^{2}}\right) ,  \notag \\
\phi (t,r) &=&\phi _{0}(t)+\sqrt{\frac{3}{4\pi G}}\frac{\alpha r_{g}H}{%
m_{\phi }u}+\mathcal{O}\left( \frac{r_{g}^{2}}{u^{2}}\right) .
\label{inflat}
\end{eqnarray}%
The second order in $r_{g}/u$ is given in appendix \ref{higher}. It
illustrates the main features of the expansion described above.

In the infinite past, $\alpha $ tends to zero, $H$ tends to $m_{\phi }/2$
and $\phi $ tends to $-\infty $, so the potential $V$ tends to $3m_{\phi
}^{2}/(32\pi G)$. Moreover, the kinetic terms of the $\phi $ Lagrangian
disappear, so the action (\ref{S}) tends to the one of gravity with a
cosmological constant $\Lambda =3m_{\phi }^{2}/4$. Correspondingly, the line
element $\mathrm{d}s^{2}$ of formula (\ref{inflat})\ tends to the line
element of a black hole with a cosmological constant in isotropic dynamic
coordinates, given by formula (\ref{dsn}). We also have%
\begin{equation*}
R_{\mu \nu \rho \sigma }R^{\mu \nu \rho \sigma }\rightarrow \frac{3}{2}%
m_{\phi }^{4}+\frac{12r_{g}^{2}}{u^{6}}\left( 1+\frac{r_{g}}{4u}\right)
^{-12},
\end{equation*}%
which agrees with what we obtain by applying the change of coordinates (\ref%
{ch1}) to (\ref{SchwC}). Using the formulas of appendix \ref{higher}, it is
possible to check these facts to the second order in $r_{g}/u$.

\subsection{Arbitrary distribution of inhomogeneities}

We can generalize the solution (\ref{inflat}) to incorporate an arbitrary
distribution of inhomogeneities, described by a potential $\Phi (x,y,z)$
with vanishing Laplacian. In isotropic dynamic coordinates, the result is 
\begin{eqnarray}
&&\mathrm{d}s^{2}=\left( \!1-\frac{2H(t)\Phi (x,y,z)}{m_{\phi }a(t)}\right)
\!\mathrm{d}t^{2}-\left( \!1+\frac{2H(t)\Phi (x,y,z)}{m_{\phi }a(t)}\right)
\!a(t)^{2}(\mathrm{d}x^{2}\!+\!\mathrm{d}y^{2}\!+\!\hspace{0.01in}\mathrm{d}%
z^{2})+\mathcal{O}(\Phi ^{2}),  \notag \\
&&\phi (t,x,y,z)=\phi _{0}(t)+\sqrt{\frac{3}{4\pi G}}\frac{\alpha (t)H(t)}{%
m_{\phi }a(t)}\Phi (x,y,z)+\mathcal{O}(\Phi ^{2}),  \label{sold}
\end{eqnarray}%
to the first order in $\Phi $, where $a(t)$, $\phi _{0}(t)$ and $\alpha (t)$
are the same as before (formulas (\ref{alfa}), (\ref{beta}) and (\ref{tau})).

\subsection{Impact on CMB anisotropies}

It is common to study the primordial inhomogeneities in the
\textquotedblleft comoving gauge\textquotedblright , where the $\phi $
fluctuation $\delta \phi $ vanishes \cite{cosmology}. Ignoring the vector
fluctuations, the metric is parametrized as 
\begin{eqnarray}
g_{\mu \nu } &=&\text{diag}(1,-a^{2},-a^{2},-a^{2})-2a^{2}\left( u\delta
_{\mu }^{1}\delta _{\nu }^{1}-u\delta _{\mu }^{2}\delta _{\nu }^{2}+v\delta
_{\mu }^{1}\delta _{\nu }^{2}+v\delta _{\mu }^{2}\delta _{\nu }^{1}\right) ,
\notag \\
&&+2\text{diag}(\tilde{\Phi},a^{2}\Psi ,a^{2}\Psi ,a^{2}\Psi )-\delta _{\mu
}^{0}\delta _{\nu }^{i}\partial _{i}B-\delta _{\mu }^{i}\delta _{\nu
}^{0}\partial _{i}B.  \label{mets}
\end{eqnarray}%
The curvature perturbation, often denoted by $\mathcal{R}$, coincides with
the field $\Psi $ of (\ref{mets}). The tensor fluctuations have $\tilde{\Phi}%
=\Psi =B=0$, while the scalar fluctuations are those with $u=v=0$.

We study the impact of the inhomogeneities contained in (\ref{sold}) on the
metric (\ref{mets}) to the leading order around the de Sitter limit. This
means that we can set $\alpha =0$ in (\ref{sold}). Then, the expression of $%
\phi $ in that formula tells us that the system is already in the comoving
gauge $\delta \phi =0$.

The metric $g_{\mu \nu }$ of (\ref{sold}) has $u=v=0$, so the
inhomogeneities incorporated in it are of the scalar type. We find $\Psi
=-H\Phi /(m_{\phi }a)=\tilde{\Phi}$, $B=0$. To the lowest order in $\alpha $%
, we can write $\tilde{\Phi}=-\dot{\Psi}/H$. This equation can also be
obtained by keeping $B$ arbitrary and integrating it out.

For definiteness, we consider a situation where the pre-existing
inhomogeneities are due to a single \textquotedblleft center of the
universe\textquotedblright\ of mass $M$. This means that we just take $\Phi
=2MG/r$, as in (\ref{inflat}). Fourier transforming the space coordinates to
momenta $\mathbf{k}$, we have ($H\simeq m_{\phi }/2$)%
\begin{equation}
\mathcal{R}_{\mathbf{k}}(\tau )=\Psi _{\mathbf{k}}(\tau )=-\frac{4\pi MG}{%
ak^{2}},  \label{pree}
\end{equation}%
where $k=|\mathbf{k}|$.

In addition to these inhomogeneities, we have the usual primordial quantum
fluctuations, 
\begin{equation}
\mathcal{R}_{\mathbf{k}}(\tau )=\psi _{\mathbf{k}}(\tau )\hat{a}_{\mathbf{k}%
}+\psi _{-\mathbf{k}}^{\ast }(\tau )\hat{a}_{-\mathbf{k}}^{\dagger }
\label{usu}
\end{equation}%
which parametrize the most general solution of the linearized equations of
motion on the\ background (\ref{sold}), with the Bunch-Davies vacuum
condition \cite{bunch}. In (\ref{usu}) $\psi _{\mathbf{k}}(\tau )$ denotes
the eigenfunctions, while $\hat{a}_{\mathbf{k}}^{\dagger }$ and $\hat{a}_{%
\mathbf{k}}$ are creation and annihilation operators, satisfying $[\hat{a}_{%
\mathbf{k}},\hat{a}_{\mathbf{k}^{\prime }}^{\dagger }]=(2\pi )^{3}\delta
^{(3)}(\mathbf{k}-\mathbf{k}^{\prime })$.

The total scalar perturbation is the sum%
\begin{equation*}
\mathcal{R}_{\mathbf{k}}(\tau )=-\frac{4\pi MG}{ak^{2}}+\psi _{\mathbf{k}%
}(\tau )\hat{a}_{\mathbf{k}}+\psi _{-\mathbf{k}}^{\ast }(\tau )\hat{a}_{-%
\mathbf{k}}^{\dagger }.
\end{equation*}%
Note that the first contribution is classical, because it is part of the
background field. The perturbation spectra are determined by the two-point
function $\langle \mathcal{R}_{\mathbf{k}}(\tau )\mathcal{R}_{\mathbf{k}%
^{\prime }}(\tau )\rangle $.

The mixed background/quantum terms of the product $\mathcal{R}_{\mathbf{k}%
}(\tau )\mathcal{R}_{\mathbf{k}^{\prime }}(\tau )$ are linear in $\hat{a}_{%
\mathbf{k}}^{\dagger }$ and $\hat{a}_{\mathbf{k}}$, so they do not
contribute to $\langle \mathcal{R}_{\mathbf{k}}(\tau )\mathcal{R}_{\mathbf{k}%
^{\prime }}(\tau )\rangle $. Since we are just interested in the lowest
orders here, we do not need to work out (\ref{usu})\ in the background (\ref%
{sold}): we can approximate the metric to the FLRW one for this purpose,
i.e., use the $G=0$ limit of (\ref{sold}). We recall that when the
corrections (\ref{pree}) are absent, one has \cite{CMBrunning}%
\begin{equation*}
\langle \mathcal{R}_{\mathbf{k}}(\tau )\mathcal{R}_{\mathbf{k}^{\prime
}}(\tau )\rangle =(2\pi )^{3}\delta ^{(3)}(\mathbf{k}+\mathbf{k}^{\prime })%
\frac{2\pi ^{2}}{k^{3}}\mathcal{P}_{\mathcal{R}},\qquad \mathcal{P}_{%
\mathcal{R}}=\frac{m_{\phi }^{2}G}{12\pi \alpha ^{2}},
\end{equation*}%
where $\mathcal{P}_{\mathcal{R}}$ is the usual power spectrum to the leading
order and $\alpha $ is the running coupling determined by the beta function (%
\ref{beta}), calculated at a conformal time $\tau $ equal to $-1/k$.

When the pre-existing inhomogeneities (\ref{pree}) are included, the result
is%
\begin{equation}
\langle \mathcal{R}_{\mathbf{k}}(\tau )\mathcal{R}_{\mathbf{k}^{\prime
}}(\tau )\rangle =\left[ \frac{(4\pi )^{2}M^{2}G^{2}}{a^{2}k^{2}k^{\prime 
\hspace{0.01in}2}}+(2\pi )^{3}\delta ^{(3)}(\mathbf{k}+\mathbf{k}^{\prime })%
\frac{\pi m_{\phi }^{2}G}{6\alpha ^{2}k^{3}}\right] (1+\mathcal{\tilde{O}}%
(\alpha ,MGk)).  \label{twopo}
\end{equation}%
The background metric we have been using is accurate to the first order in $%
MG/r$, so the right-hand side is right up to corrections of orders $MGk$, $%
MGk^{\prime }$, as well as further corrections like the ones described in
appendix \ref{higher}. We incorporate all of them into the symbol $\mathcal{%
\tilde{O}}(\alpha ,MGk)$.

The background and quantum contributions appearing on the right-hand side of
(\ref{twopo}) are very different, and it is not straightforward to compare
them. To capture the full complexity of the primordial fluctuations in
models with pre-existing inhomogeneities, additional measurements beyond the
common power spectrum are probably needed. What we can do right away is
compare averages that place the two terms somehow on the same footing.

We take the common \textquotedblleft pivot\textquotedblright\ scale $k_{\ast
}=0.05$ Mpc$^{-1}$ and integrate on the range of momenta $C=\{k$ such that $%
k_{\min }\equiv 10^{-4}$ Mpc$^{-1}\leqslant k\leqslant 1$ Mpc$^{-1}=k_{\max
}\}$, which is the most studied observationally. Using the data of \cite%
{Planck18}, we find%
\begin{equation*}
\alpha _{\ast }=0.0087\pm 0.0010,\qquad m_{\phi }=(2.99\pm 0.37)\cdot 10^{13}%
\text{GeV.}
\end{equation*}%
Approximating $\alpha $ to its pivot value $\alpha _{\ast }$, we get%
\begin{equation}
\int_{C}\frac{\mathrm{d}^{3}\mathbf{k}}{(2\pi )^{3}}\int_{C}\frac{\mathrm{d}%
^{3}\mathbf{k}^{\prime }}{(2\pi )^{3}}\langle \mathcal{R}_{\mathbf{k}}(\tau )%
\mathcal{R}_{\mathbf{k}^{\prime }}(\tau )\rangle \simeq \frac{4M^{2}G^{2}}{%
\pi ^{2}a^{2}}(k_{\max }-k_{\min })^{2}+\frac{m_{\phi }^{2}G}{12\pi \alpha
_{\ast }^{2}}\ln \frac{k_{\max }}{k_{\min }}.  \label{ave}
\end{equation}

This result is valid up to the horizon re-entry. It must then be evolved up
to the last scattering surface by means of appropriate transfer functions,
in order to relate it to the CMB observations \cite{cosmology}. This is not
an easy task.

Observe that the terms appearing on the right-hand side of (\ref{ave}) are
functions of $k$, apart from a factor $1/a^{2}$ multiplying the background
contribution. That factor must evolve into $1/a^{2}\simeq 10^{6}$, that is
to say, the same factor calculated at the time of last scattering (where $%
a\simeq 10^{-3}$).

It remains to evolve the two functions of $k$. Since we are dealing with
averages on $k$, we expect that they will not be impacted in dramatically
different ways. Then we can argue that the contributions of the pre-existing
inhomogeneities are comparable to the usual anisotropies when the black-hole
mass $M$ is%
\begin{equation}
M=\frac{m_{\phi }a}{4\alpha _{\ast }(k_{\max }-k_{\min })}\sqrt{\frac{\pi }{%
3G}\ln \frac{k_{\max }}{k_{\min }}}\simeq 4\cdot 10^{12}M_{\odot }\simeq
10^{43}\text{kg}.  \label{massa}
\end{equation}%
This $M$ is ten thousand times heavier than the heaviest black holes known
today.

The outcome is that there is room for interesting perennial inhomogeneities
in the universe without contradicting present knowledge.

\section{The fate of neutron stars and white dwarfs}

\label{neutron}\setcounter{equation}{0}

In this section we investigate the effects of the expansion of the universe,
due to the cosmological constant $\Lambda $, on the equilibrium
configuration of a neutron star, a white dwarf, or, more generally, a system
where the fermion degeneracy pressure opposes the gravitational force. For
simplicity, we work at zero temperature, which is, strictly speaking, the
case of a black dwarf. We want to determine whether the star remains in
equilibrium with respect to the event horizon ($ra=$ constant), or
contracts, or expands.

Mimicking (\ref{dsn}), we assume an isotropic metric 
\begin{equation}
\mathrm{d}s^{2}=g_{00}(u)\mathrm{d}t^{2}-a(t)^{2}g_{r}(u)(\mathrm{d}x^{2}+%
\mathrm{d}y^{2}+\hspace{0.01in}\mathrm{d}z^{2}),  \label{dsndw}
\end{equation}%
where $g_{00}$ and $g_{r}$ are unknown functions of $u=ra(t)$, and $a(t)$ is
the one of (\ref{at}), with $H=\sqrt{\Lambda /3}$. The Lagrangian of a
particle of mass $m$ and its momentum read%
\begin{equation}
L=-m\sqrt{g_{00}-g_{r}a^{2}v^{2}},\qquad \mathbf{p}=\frac{\partial L}{%
\partial \mathbf{v}}=-m^{2}a^{2}g_{r}\frac{\mathbf{v}}{L},  \label{ll}
\end{equation}%
where $\mathbf{v}=\mathrm{d}\mathbf{r}/\mathrm{d}t$ is the velocity and $v=|%
\mathbf{v}|$. The equations of motion are%
\begin{equation}
\frac{\mathrm{d}\mathbf{p}}{\mathrm{d}t}=\frac{\partial L}{\partial \mathbf{r%
}}=\frac{am^{2}\mathbf{r}}{2Lr}(g_{00}^{\prime }-g_{r}^{\prime
}a^{2}v^{2})\equiv \mathbf{f}_{G}.  \label{force}
\end{equation}

To inquire whether the star expands or contracts with respect to the event
horizon, we need to generalize the equations in several respects. First, we
have to account for a (spherically symmetric) distribution of matter, rather
than a pointlike mass placed at the center. This means that we must work
with the equations of fluid dynamics in general relativity, derived in
appendix \ref{fluidGR}. Second, we need to include the effects of the
degeneracy pressure, due to the Pauli exclusion principle, and use the
equations of state of degenerate fermions (derived in appendix \ref%
{degeneracy}). Third, at some point we need to switch to coordinates $%
\mathbf{u}=\mathbf{r}a$, where the event horizon is stationary.

We start from the generalization of (\ref{force}) to a fluid. The equations
of motion (\ref{flM}) of fluid dynamics, which we repeat here for
convenience, are%
\begin{equation}
(\varepsilon +P)U^{\nu }D_{\nu }p^{\mu }=m(g^{\mu \nu }-U^{\nu }U^{\mu
})D_{\nu }P,  \label{flneu}
\end{equation}%
where $p^{\mu }=mU^{\mu }$ is the four momentum, $P=P(t,\mathbf{r})$ is the
pressure and $\varepsilon =\varepsilon (t,\mathbf{r})$ is the energy density
(as defined in appendix \ref{fluidGR}). The equation of state $%
P=P(\varepsilon )$ obeyed by an ideal Fermi fluid at zero temperature is
given in appendix \ref{degeneracy}. In most arguments we do not need to
specify it, but just assume that it exists and is such that $P(\varepsilon
)/\varepsilon $ tends to zero for $\varepsilon \rightarrow 0$.

More explicitly, lowering the index $\mu $ in (\ref{flneu}) and specializing
to space indices $\mu =i$, (\ref{flneu}) gives equation (\ref{fleqmP}),
which reads, in the case we are considering here,%
\begin{equation}
\frac{\partial \mathbf{p}}{\partial t}+(\mathbf{v}\hspace{0.01in}\cdot %
\bm{\nabla })\mathbf{p}=\mathbf{f}_{G}+\frac{L}{\varepsilon +P}\bm{\nabla }P-%
\frac{\mathbf{p}}{\varepsilon +P}\left[ \frac{\partial P}{\partial t}+(%
\mathbf{v}\hspace{0.01in}\cdot \bm{\nabla })P\right] \equiv \mathbf{f}_{G}+%
\mathbf{f}_{P}.  \label{fleq}
\end{equation}

For a specific fluid, $P$ and $\varepsilon $ are scalars, and functions of
another scalar density $\rho $, which is related by formula (\ref{rot}) to
the density of mass $\rho _{m}(t,\mathbf{r})=\mathrm{d}m/\mathrm{d}^{3}%
\mathbf{r}=m\rho _{n}(t,\mathbf{r})$, where $\rho _{n}(t,\mathbf{r})$ is the
number of particles per unit volume in a given system of coordinates.
Constraints on $\rho _{m}$ are the total mass 
\begin{equation}
M=4\pi \int_{0}^{\infty }r^{2}\mathrm{d}r\rho _{m}(t,\mathbf{r})
\label{norma}
\end{equation}%
of the fluid distribution and the continuity equation (\ref{CC}), which we
rewrite here as well: 
\begin{equation}
\frac{\partial \rho _{m}}{\partial t}+\bm{\nabla }\cdot (\rho _{m}\mathbf{v}%
)=0.  \label{conta}
\end{equation}

Finally, the metric must obey the Einstein equations (\ref{E})\ with the
energy-momentum tensor $T^{\mu \nu }=(\varepsilon +P)U^{\mu }U^{\nu
}-Pg^{\mu \nu }$. The complete set of equations is summarized in formula (%
\ref{fluiddynGR}).

\subsection{Static coordinates\label{staticco}}

It is convenient to introduce \textquotedblleft static\textquotedblright\
coordinates $\mathbf{u}=\mathbf{r}a$ (as opposed to the \textquotedblleft
dynamic\textquotedblright\ coordinates $\mathbf{r}$), where the event
horizon does not depend on time. This simplifies various expressions, since $%
g_{00}$ and $g_{r}$ are just functions of $u=ra$. If $\mathbf{P}$ denotes
the momentum associated with $\mathbf{u}$ (not to be confused with the
pressure $P$), the switch $(\mathbf{r},\mathbf{p})\leftrightarrow (\mathbf{u}%
,\mathbf{P})$ is a canonical transformation. Due to this, certain
operations, like converting the quantization rules from one choice of
coordinates to the other, are straightforward. Note that the definition of
time remains the same in the switch.

In static coordinates, the Lagrangian, momentum and equation of motion of a
single particle read 
\begin{eqnarray}
L &=&-m\sqrt{g_{00}-g_{r}(\mathbf{\dot{u}}-H\mathbf{u})^{2}},\qquad \mathbf{P%
}=\frac{\partial L}{\partial \mathbf{\dot{u}}}=\frac{\mathbf{p}}{a}=-\frac{%
m^{2}}{L}g_{r}(\mathbf{\dot{u}}-H\mathbf{u}),  \notag \\
\frac{\mathrm{d}\mathbf{P}}{\mathrm{d}t} &=&\frac{\partial L}{\partial 
\mathbf{u}}=-H\mathbf{P}+\frac{m^{2}\mathbf{u}}{2uL}(g_{00}^{\prime
}-g_{r}^{\prime }(\mathbf{\dot{u}}-H\mathbf{u})^{2})\equiv -H\mathbf{P}+%
\mathbf{F}_{G}.  \label{uvar}
\end{eqnarray}

Before proceeding, we pause a moment to identify the circular orbits, which
are those with $u=$ constant, $\mathbf{\dot{u}}\cdot \mathbf{u}=0$.
Contracting the equation of motion appearing in the second line of (\ref%
{uvar}) with $\mathbf{u}$, and using $\mathbf{\dot{u}}\cdot \mathbf{\dot{u}}+%
\mathbf{\ddot{u}}\cdot \mathbf{u}=0$, it is easy to prove that $\mathbf{%
\ddot{u}}\cdot \mathbf{u}$ is constant. Hence, $\mathbf{\dot{u}}\cdot 
\mathbf{\dot{u}}$ is also constant, and $\mathbf{\dot{u}}\cdot \mathbf{\ddot{%
u}}=0$. This means that we can write $\mathbf{\ddot{u}=-}\omega ^{2}\mathbf{u%
}$, where $\omega $ is the angular velocity, after which we infer $|\mathbf{%
\dot{u}}|=\omega u$. Finally, the equation of motion gives the relation%
\begin{equation}
u(2g_{r}+ug_{r}^{\prime })(\omega ^{2}+H^{2})=g_{00}^{\prime },
\label{circular}
\end{equation}%
which determines the right $\omega $ for every distance $u$ from the center.
In the Newton approximation of the metric (\ref{dsn}), where%
\begin{equation}
g_{00}\simeq 1-\frac{2MG}{u},\qquad g_{r}\simeq 1+\frac{2MG}{u},  \label{N}
\end{equation}%
the condition (\ref{circular})\ becomes%
\begin{equation}
\omega ^{2}+H^{2}\simeq \frac{MG}{u^{3}}.  \label{orbit}
\end{equation}

Formulas (\ref{circular}) and (\ref{orbit}) show that the expansion rate $H$
and the angular velocity $\omega $ are on the same footing, i.e., the
\textquotedblleft force\textquotedblright\ due to the expansion of the
universe is similar to a centrifugal force. Moreover, the orbits exist only
within a certain maximum distance from the center, which is equal to $\sqrt[3%
]{MG/H^{2}}$ in the approximation (\ref{orbit}). Beyond this region, the
attractive force of gravity becomes too weak to counterbalance the
centrifugal force of expansion.

The result is compatible with the one of ref. \cite{Price}, that is to say,
weakly coupled systems are comoving while strongly coupled ones resist the
cosmic expansion.

We have just learned that when the orbiting system exists, it does not
expand. This suggests that fluids might also admit equilibrium
configurations that deplete the expansion of the universe, as long as the
expansion rate is not excessive. In the rest of this section we show that it
is indeed so.

It is convenient to convert the fluid equations (\ref{fleq}) to the
variables $\mathbf{u}=a\mathbf{r}$. Noting that $\mathbf{p}(t,\mathbf{r})=a%
\mathbf{P}(t,\mathbf{u})$ and $\mathbf{v=(\dot{u}}-H\mathbf{u})/a$, we have%
\begin{equation}
\frac{\partial \mathbf{p}}{\partial t}+(\mathbf{v}\hspace{0.01in}\cdot %
\bm{\nabla })\mathbf{p}=a\left( \frac{\partial \mathbf{P}}{\partial t}+(%
\mathbf{\dot{u}}\hspace{0.01in}\cdot \bm{\nabla })\mathbf{P}+H\mathbf{P}%
\right) ,  \label{depa}
\end{equation}%
where $\bm{\nabla }$ is $\mathbf{\partial /\partial r}$ on $\mathbf{p}$ and $%
\mathbf{\partial /\partial u}$ on $\mathbf{P}$. Equation (\ref{fleq}) turns
into%
\begin{eqnarray}
\frac{\partial \mathbf{P}}{\partial t}+(\mathbf{\dot{u}}\hspace{0.01in}\cdot %
\bm{\nabla })\mathbf{P} &=&-H\mathbf{P}+\frac{m^{2}\mathbf{u}}{2uL}%
(g_{00}^{\prime }-g_{r}^{\prime }(\mathbf{\dot{u}}-H\mathbf{u})^{2})  \notag
\\
&&+\frac{L}{\bar{\varepsilon}+\bar{P}}\bm{\nabla }\bar{P}-\frac{\mathbf{P}}{%
\bar{\varepsilon}+\bar{P}}\left[ \frac{\partial \bar{P}}{\partial t}+(%
\mathbf{\dot{u}}\hspace{0.01in}\cdot \bm{\nabla })\bar{P}\right] \equiv 
\mathbf{F}_{c}+\mathbf{F}_{G}+\mathbf{F}_{P}\mathbf{,\qquad }  \label{flequ}
\end{eqnarray}%
where we have defined $\varepsilon (t,\mathbf{r})=\bar{\varepsilon}(t,%
\mathbf{u})$ and $P(t,\mathbf{r})=\bar{P}(t,\mathbf{u})$, so $\bm{\nabla }%
\bar{P}=\partial \bar{P}\mathbf{/\partial u}$.

The expression between the equal and equivalence signs in equation (\ref%
{flequ}) is the sum of the \textquotedblleft centrifugal
force\textquotedblright\ $\mathbf{F}_{c}=\mathbf{-}H\mathbf{P}$ due to the
expansion of the universe, plus the gravitational force $\mathbf{F}_{G}$,
plus the pressure force $\mathbf{F}_{P}$. The last two coincide with the
forces $\mathbf{f}_{G}$ and $\mathbf{f}_{P}$ of formulas (\ref{force}), (\ref%
{fleq}) and (\ref{forces}), divided by $a$.

Further defining $\rho _{m}(t,\mathbf{r})=a^{3}\bar{\rho}_{m}(t,\mathbf{u})$%
, the continuity equation (\ref{conta}) gives 
\begin{equation}
\frac{\partial \bar{\rho}_{m}}{\partial t}+\bm{\nabla }\cdot (\bar{\rho}_{m}%
\mathbf{\dot{u}})=0.  \label{contau}
\end{equation}

\subsection{Solution of the equations}

We search for (spherically symmetric) hydrodynamic equilibrium
configurations in static coordinates. That is to say, we set $\mathbf{\dot{u}%
}=0$, which gives $\mathbf{v=}-H\mathbf{r}$. Formula (\ref{contau}) tells us
that $\bar{\rho}_{m}$ does not depend explicitly on time, so from now on we
write $\bar{\rho}_{m}=\bar{\rho}_{m}(u)$. Formula (\ref{rot}) gives%
\begin{equation}
\rho =\frac{\bar{\rho}_{m}(u)\sqrt{g_{00}-g_{r}u^{2}H^{2}}}{g_{r}\sqrt{%
g_{00}g_{r}}},  \label{rotu}
\end{equation}%
which also depends just on $u$. This property extends to $P$ and $%
\varepsilon $, which are functions of $\rho $ by the equations of state and
the identity (\ref{rotilde}). Thus, we write $\rho =\bar{\rho}(u)$, $P=\bar{P%
}(u)$ and $\varepsilon =\bar{\varepsilon}(u)$. Finally, formula (\ref{uvar})
shows that the momenta 
\begin{equation}
\mathbf{P}=-\frac{mHg_{r}\mathbf{u}}{\sqrt{g_{00}-g_{r}u^{2}H^{2}}}
\label{Pu}
\end{equation}%
depend just on $\mathbf{u}$.

Collecting these pieces of information, the fluid equation (\ref{flequ})
gives%
\begin{equation}
\frac{1}{\bar{\varepsilon}+\bar{P}}\frac{\mathrm{d}\bar{P}}{\mathrm{d}u}=-%
\frac{g_{00}^{\prime }-uH^{2}(2g_{r}+ug_{r}^{\prime })}{%
2(g_{00}-g_{r}u^{2}H^{2})}=-\frac{\mathrm{d}}{\mathrm{d}u}\ln \sqrt{%
g_{00}-g_{r}u^{2}H^{2}}.  \label{eq2}
\end{equation}%
Assuming the equation of state $P=P(\varepsilon )$ and using (\ref{important}%
) and (\ref{rotilde}) to express $\rho $ as a function of $\varepsilon $ as
well, we can integrate (\ref{eq2}) to find 
\begin{equation}
\int^{\bar{\varepsilon}(u)}\frac{\mathrm{d}\varepsilon }{\rho (\varepsilon )}%
\frac{\mathrm{d}\rho }{\mathrm{d}\varepsilon }\frac{\mathrm{d}P}{\mathrm{d}%
\varepsilon }=C-\ln \sqrt{g_{00}-g_{r}u^{2}H^{2}},  \label{solP}
\end{equation}%
where $C$ is the integration constant. This formula gives $\bar{\varepsilon}%
(u)$ implicitly, wherefrom $\bar{P}$, $\bar{\rho}$ and $\bar{\rho}_{m}$
follow, hence $\varepsilon $, $P$, $\rho $ and $\rho _{m}$.

In the case of an ideal degenerate Fermi fluid, the primitive on the
left-hand side is reported in formula (\ref{primitive}) as a function of $%
\rho $. We find%
\begin{equation}
\rho (u)=\left( \frac{\hat{\rho}^{2/3}}{g_{00}(u)-g_{r}(u)u^{2}H^{2}}-\rho
_{\ast }^{2/3}\right) ^{3/2},\qquad \rho _{\ast }\equiv \frac{8\pi m^{4}}{%
3h^{3}},  \label{ru}
\end{equation}%
where $\hat{\rho}$ is a constant.

We can distinguish two regions: $u\leqslant u_{\max }$ and $u>u_{\max }$,
where $u_{\max }$ denotes the radius of the star. Clearly, we must have $%
u_{\max }<u_{\text{hor}}$, where $u_{\text{hor}}$ denotes the event horizon,
which is where $g_{00}(u_{\text{hor}})=g_{r}(u_{\text{hor}})u_{\text{hor}%
}^{2}H^{2}$. Since $\rho $ is proportional to $\rho _{m}$ by (\ref{rot}), or
(\ref{rotu}), we must have $\rho =0$ outside the star. This fixes the
constant $\hat{\rho}$. The result is%
\begin{equation}
\left\{ 
\begin{tabular}{ll}
$\rho (u)=\rho _{\ast }\left( \dfrac{g_{00}(u_{\max })-g_{r}(u_{\max
})u_{\max }^{2}H^{2}}{g_{00}(u)-g_{r}(u)u^{2}H^{2}}-1\right) ^{3/2}$ & $%
\qquad \text{for }u\leqslant u_{\max },$ \\ 
$\rho (u)=0$ & $\qquad \text{for }u>u_{\max }.$%
\end{tabular}%
\right.  \label{dens}
\end{equation}%
The normalization condition (\ref{norma}), which becomes%
\begin{equation}
M=4\pi \int_{0}^{u_{\max }}u^{2}\mathrm{d}u\bar{\rho}_{m}(u),  \label{normaH}
\end{equation}%
can be used to trade $u_{\max }$ for $M$, or vice versa.

Equipped with the result (\ref{dens}), formula (\ref{rot}) gives $\rho _{m}$%
, the identity (\ref{rotilde}) gives $\varepsilon $ and the equation of
state $P=P\left( \varepsilon \right) $ gives the pressure.

Outside the star, we have $\rho _{m}=\rho =0$. Having assumed that $%
P(\varepsilon )/\varepsilon $ tends to zero for $\varepsilon \rightarrow 0$,
which is true for an ideal degenerate Fermi fluid (as shown right below
equation (\ref{state})), formula \ref{rotilde} implies that $\varepsilon $
also vanishes when $\rho =0$, so $P$ vanishes as well. By the same
arguments, $\rho $, $\rho _{m}$, $\varepsilon $ and $P$ tend to zero when $u$
tends to $u_{\max }$ from the inside. Finally, formula (\ref{eq2}) implies
that the gradient of the pressure (which encodes the force $\mathbf{F}_{P}$
due to it) tends to zero as well. This means that the star does not lose
particles from its exterior border.

It remains to determine the metric. It is easy to check that the Einstein
equations (\ref{E}) lose the powers of $a$ and provide the remaining
equations for $g_{00}$ and $g_{r}$. In the end, we have five unknown
functions of $u$, which are $g_{00}$, $g_{r}$, $\rho $, $\varepsilon $ and $%
P $, three independent differential equations (two from (\ref{E}) plus (\ref%
{eq2}) from (\ref{flequ})), an equation of state relating $\varepsilon $ and 
$P$, and a universal relation (\ref{rotilde}) between $\varepsilon $ and $%
\rho $. With the boundary condition $\rho (u_{\max })=0$, the system admits
a solution under a certain condition that we specify in a moment.

The functions $g_{00}$ and $g_{r}$ can be worked out as expansions in powers
of $G$. The results to order one are%
\begin{eqnarray}
g_{00}(u) &=&1-\frac{4GH^{2}}{u}\mu _{2}(\nu _{1},u)-\frac{2G}{u}\nu _{2}(u)+%
\frac{2G}{3}\left( 3\nu _{1}(u)-2G\nu _{1}(u_{\max })\right) ,  \notag \\
g_{r}(u) &=&2-g_{00}(u)-4GH^{2}\mu _{1}(\nu _{1},u),\qquad  \label{fo}
\end{eqnarray}%
both inside and outside $u_{\max }$, having defined the moments%
\begin{equation}
\nu _{k}(u)=4\pi \int_{0}^{u}\frac{w^{k}\tilde{\varepsilon}(w)\mathrm{d}w}{%
1-w^{2}H^{2}},\qquad \mu _{k}(f,u)=\int_{0}^{u}w^{k}f(w)\mathrm{d}w,
\label{foa}
\end{equation}%
where $\tilde{\varepsilon}$ is the sum $\varepsilon +P$, which can be
derived recursively from (\ref{dens}), (\ref{rotilde}) and the equation of
state (\ref{state}). Clearly, we must have $u_{\max }<1/H$ in the expansion.

The system does not admit a solution for an arbitrary mass $M$, or an
arbitrary radius $u_{\max }<1/H$. A simple way to appreciate this fact is by
comparing the cases $G=0$ and the nonrelativistic limit at $H=0$.

When the Newton constant $G$ is switched off, the Einstein equations (\ref{E}%
) are solved by the FLRW\ metric ($g_{00}=g_{r}=1$), so (\ref{dens}) gives%
\begin{equation}
\rho (u)=\rho _{\ast }H^{3}\left( \dfrac{u^{2}-u_{\max }^{2}}{1-u^{2}H^{2}}%
\right) ^{3/2}  \label{G=0}
\end{equation}%
for $u<u_{\max }<1/H$. The argument of the fractional power is negative, so
the solution is not acceptable, or we can say that it forces $u_{\max }=0$.
The only possibility to have something mathematically meaningful at $G=0$ is
to renounce $\rho (u_{\max })=0$ (the condition that the star has an
exterior boundary) and take $u_{\max }=1/H$. Switching back to (\ref{ru}),
we find the density%
\begin{equation}
\rho (u)=\left( \frac{\hat{\rho}^{2/3}}{1-u^{2}H^{2}}-\rho _{\ast
}^{2/3}\right) ^{3/2},  \label{nonrela}
\end{equation}%
where the constant $\hat{\rho}$ remains free, provided it is large enough.
Formula (\ref{norma}) implies that the total mass $M$ is infinite.

The density (\ref{nonrela}) grows together with the distance from the
center, and tends to infinity when $u$ approaches $1/H$. This solution is
not physically realistic, but serves to illustrate what is required to
compensate for the centrifugal force $\mathbf{F}_{c}=-H\mathbf{P}$ due to
the expansion of the universe in the absence of the gravitational force $%
\mathbf{F}_{G}$.

In the nonrelativistic limit (\ref{nrstate}) at $H=0$, formula (\ref{dens})
and (\ref{fo}) give back the known results \cite{Stars}, which describe the
compensation between gravitational attraction and fermion degeneracy
pressure in objects such as neutron stars and white dwarfs. Precisely,
noting that $u=r$, $\tilde{\varepsilon}\simeq \varepsilon \simeq \rho $ and,
to the lowest order in $G$, $\rho \simeq \rho _{m}$, we find 
\begin{equation}
g_{00}(r)\simeq 1+2G\int_{0}^{r}\frac{\mathrm{d}w}{w^{2}}M(w)-\frac{2G}{3}%
\int_{0}^{r_{\max }}\frac{\mathrm{d}w}{w}\frac{\mathrm{d}M(w)}{\mathrm{d}w}%
,\qquad g_{r}\simeq 2-g_{00},\text{ }  \label{usual}
\end{equation}%
and%
\begin{equation}
P(r)\simeq G\int_{r}^{r_{\max }}\frac{\mathrm{d}w}{w^{2}}\hspace{0.01in}\rho
_{m}(w)M(w)\qquad (r<r_{\max }),  \label{usualpress}
\end{equation}%
where%
\begin{equation*}
M(r)=4\pi \int_{0}^{r}w^{2}\rho _{m}(w)\mathrm{d}w
\end{equation*}%
is the mass contained within $r$. The metric (\ref{usual}) is correct
outside the star as well as inside, while the pressure (\ref{usualpress}) is
zero outside. The solution makes sense, because the argument of the
fractional power in formula (\ref{dens}) is positive. Actually, that formula
gives the integral equation%
\begin{equation*}
\left( \frac{\rho _{m}(r)}{\rho _{\ast }}\right) ^{2/3}\simeq 8\pi
G\int_{r}^{r_{\max }}\frac{\mathrm{d}w}{w^{2}}\int_{0}^{w}z^{2}\rho _{m}(z)%
\mathrm{d}z,
\end{equation*}%
which, together with the conditions $\rho _{m}(r_{\max })=0$, determines $%
\rho _{m}(r)$. This part of the problem can be solved numerically. Here we
content ourselves with the behavior of $\rho _{m}(r)$ for $r\lesssim r_{\max
}$, which is%
\begin{equation*}
\rho _{m}(r)\simeq \rho _{\ast }\left[ \frac{2GM}{r_{\max }}\left( 1-\frac{r%
}{r_{\max }}\right) \right] ^{3/2}\qquad (r\lesssim r_{\max }).
\end{equation*}

We briefly comment on the nonrelativistic limit at $G=0$, $H\neq 0$, where (%
\ref{solP}) and (\ref{nrstate}) give 
\begin{equation}
\bar{\rho}_{m}(u)=\rho _{\ast }\frac{\left[ 2C-\ln (1-u^{2}H^{2})\right]
^{3/2}}{\sqrt{1-u^{2}H^{2}}},  \label{solnr}
\end{equation}%
the constant $C$ being fixed by (\ref{normaH}). In this case, $M$ is finite
and must exceed a minimum value, obtained by choosing $C=0$ in (\ref{solnr}%
). The reason of this apparent contradiction between the general case (\ref%
{nonrela}), where $M$ is infinite, and the nonrelativistic approximation (%
\ref{solnr}), where $M$ is finite, is that the configuration we are
considering has velocity $\mathbf{v=}-H\mathbf{r}$, so it is impossible to
have a meaningful nonrelativistic approximation everywhere: for $u\simeq 1/H$
the particles necessarily reach the velocity of light $v=1/a$. Apart from
this, the approximate configuration (\ref{solnr}) exhibits the same
unrealistic properties of the solution (\ref{nonrela}).

The existence of opposite (unrealistic/realistic) cases $G=0$, $H\neq 0$ and 
$G\neq 0$, $H=0$, shows that a certain condition must be fulfilled to make
the solution encoded in (\ref{dens}) and (\ref{fo}) acceptable.
Specifically, the argument of the fractional power of (\ref{dens}) must be
nonnegative for every $u<u_{\max }$. In general, it is involved to unfold
this requirement. Yet, once we insert (\ref{fo}) into (\ref{dens}), we can
simplify the condition, in the nonrelativistic limit and to the first order
in $G$, by replacing $\tilde{\varepsilon}$ with the average density $M/V$,
where $V=4\pi u_{\max }^{3}/3$ is the volume of the star. The result is that
the solution is acceptable if%
\begin{equation}
u_{\max }^{2}H^{2}\lesssim \frac{GM}{u_{\max }}\ll 1.  \label{condition}
\end{equation}%
The right inequality is the condition to have a meaningful expansion in
powers of the Newton constant. The left inequality says that the expansion
of the universe cannot be too rapid.

Precisely, formula (\ref{Pu}) gives $\mathbf{F}_{c}\simeq mH^{2}\mathbf{u}$
at $G=0$ (assuming $u_{\max }^{2}H^{2}\ll 1$), so $|\mathbf{F}_{c}|\sim
mH^{2}u_{\max }$ at the border of the star. The left inequality of (\ref%
{condition}) says that the star cannot exist unless the centrifugal force $%
\mathbf{F}_{c}$ due to expansion is smaller than the gravitational force $%
\mathbf{F}_{G}$ at the border ($|\mathbf{F}_{G}|\sim GmM/u_{\max }^{2}$). If
the inequality is violated, we get into the phase where the solution does
not make sense mathematically, as in (\ref{G=0}).

With the present value of the Hubble parameter, ordinary stars, as well as
elementary or fundamental particles, fulfill the left condition (\ref%
{condition}) by tens of orders of magnitude. The sun turned into white dwarf
has $r\equiv GM/(u_{\max }^{3}H^{2})\simeq 10^{34}$, while the proton has $%
r\simeq 10^{43}$. Given that $H$ has not changed significantly since the
time of last scattering, the same conclusion has been valid since then, long
before the formation of stars. However, the condition is violated ($r\simeq
10^{-75}$ and $10^{-66}$, respectively) by the value $\simeq m_{\phi }/2$ of 
$H$ during inflation. This means that there has been a time when the
hydrostatic equilibrium was impossible, because the centrifugal force due to
the expansion of the universe was superior to the gravitational force. We
could say that no elementary particles existed, at that time, because they
would have been \textquotedblleft dismembered\textquotedblright\ by the
expansion.

Also note that objects located far enough from the center of attraction
escape due to expansion, because the gravitational force, which decreases
with distance, is no longer able to balance the centrifugal force of
expansion, which increases with distance.

Finally, it is worth to stress that the realistic, physical solution at $G=0$
is not (\ref{G=0}), or (\ref{solnr}), but the homogeneous one, where the
metric is still the FLRW\ one ($g_{00}=g_{r}=1$), but the particles are at
rest in the dynamic variables $\mathbf{r}$ ($\mathbf{v}=0$). Then the
density $\rho _{m}$ is constant by the continuity equation (\ref{conta}),
the scalar $\rho =\rho _{m}/a^{3}$ depends only on $t$ by (\ref{rot}), $P$
and $\varepsilon $ depend only on $t$ as well, by the equation of state and (%
\ref{rotilde}). Finally, $\mathbf{p}\propto \mathbf{v}=0$ by (\ref{ll}) and $%
\mathbf{f}_{G}=0$ by (\ref{force}), so the fluid equation (\ref{fleq}) is
satisfied. By homogeneity, the total mass $M$ is infinite.

Explicitly, the solution 
\begin{equation*}
\mathbf{r}=\mathbf{r}_{0}-\frac{\mathbf{p}_{0}}{aHp_{0}^{2}}\sqrt{%
p_{0}^{2}+m^{2}a^{2}}
\end{equation*}%
of the equations of motion (\ref{force}) of a single particle in the $G=0$
limit, where $\mathbf{r}_{0}$ and $\mathbf{p}_{0}$ are the integration
constants, shows that the state of equilibrium ($t\rightarrow \infty $) is
the one with $\mathbf{r}=$ constant $=\mathbf{r}_{0}-m\mathbf{p}%
_{0}/(Hp_{0}^{2})$. Clearly, this state of equilibrium has little to do with
a star.

\section{The fate of the universe}

\label{fate}\setcounter{equation}{0}

In about five billion years, the sun will expand into a red giant. Over the
course of millions of years after that, it will shed its outer layers and
transform into a white dwarf. At that point, the gravitational force will be
balanced by the electron degeneracy pressure, preventing further
gravitational collapse. In the rest of time, the white dwarf will slowly
lose its heat by radiating. There is theoretical speculation that, in a
hugely extended timeframe, the white dwarf will eventually cool down
completely and become a \textquotedblleft black dwarf\textquotedblright . In
that process, the balance between gravity and the fermion degeneracy
pressure\ will remain in place.

The black dwarf is considered to be stable. The results of the previous
section confirm that it is stable even when we take into account the
expansion of the universe.

The event horizon is $u=ra=$ constant. The Schwartzschild radius is also
given by $u=ra=$ constant, as emphasized by (\ref{dsn}), (\ref{ch1})\ and (%
\ref{statiso}). To simplify as much as possible, we face three
possibilities: $a$) if a celestial body expands in the $\mathbf{u}$
variables, its constituents eventually reach the state of total dilution; $b$%
) if it stays in equilibrium, it resists the veer toward that fate; $c$) if
it contracts, it collapses to form a black hole.

Eventually, the black hole evaporates, emitting particles and radiation. The
emitted particles are expected to move away and disperse into space,
traveling indefinitely and becoming increasingly diluted. This way, more and
more particles eventually reach the state of total dilution. So, the cases $%
a $) and $c$) lead to the same outcome. Only the intermediate situation of
equilibrium $b$) can resist the drift toward the ultimate fate.

In some sense, black-hole evaporation is a way to counteract the
gravitational attraction and make the expansion of the universe prevail. In
addition, an event horizon may not be necessary to have emission of
radiation \cite{Wondrak}, and evaporation. A system could produce pairs and
lose energy through a gravitational analogue \cite{gravSHES,Wilczek,Wondrak}
of the Schwinger pair production mechanism \cite{SHES}. If so, the
equilibrium $b$) would just be temporary. Nature would host a more
\textquotedblleft democratic\textquotedblright\ process (in the sense that
it would not be just the prerogative of a privileged class, such as black
holes) to counteract gravitational attraction and sustain expansion over
time, toward the final state of total dilution. This scenario aligns with
the idea that quanta do not favor stability but uncertainty, providing
avenues for escape. At any rate, we root for total dilution here, although
further study is needed before reaching a verdict on the fate of the
universe.

Because mechanisms like the ones just mentioned must be advocated, the
evolution toward the candidate ultimate state may take an inordinate amount
of time. For reference, Cygnus X-1, a \textquotedblleft
light\textquotedblright\ black hole with mass $M$ equal to 15 solar masses $%
M_{\odot }$, takes about $7\cdot 10^{70}$ years to completely evaporate%
\footnote{%
The general formula for the Hawking evaporation time is $t_{H}\simeq 2\cdot
10^{67}(M/M_{\odot })^{3}$ years.}. While time spans of this magnitude are
exceptionally lengthy, physicists are undeterred in the exploration of the
phenomena that require them.

Once the state of total dilution is reached, say at time $t_{\text{dis}}$,
the isolated particles are not actually particles, but wave functions, bound
to remain so forever, because they cannot collapse at later times. Particles
may receive signals from other particles after $t_{\text{dis}}$, coming from
their past histories. However, those signals are just radiation. Even if
particle A starts a journey toward particle B before $t_{\text{dis}}$, it is
unable to reach B at $t>t_{\text{dis}}$, because it would have to overcome
the dilution occurred at $t_{\text{dis}}$. We do not know of wave-function
collapses triggered by radiation only.

The \textquotedblleft irony\textquotedblright\ is that virtuality, which has
no classical counterpart and is normally confined to microscopic physics,
may be destined to ultimately embrace the vast immensity of the universe. In
fact, the investigation carried out in this paper was inspired by this very
idea. In quantum mechanics, virtuality plays a key role, by mathematically
filling the gap between two subsequent measurements on a system.
Entanglement is a striking manifestation of the virtual nature of quantum
states. The \textquotedblleft predominance\textquotedblright\ of virtuality
over reality is also apparent in quantum field theory, where a propagator is
almost everywhere virtual, the sole exception being its (relatively tiny)
on-shell contribution. Another place where virtuality plays a crucial role
is quantum gravity, where it is possible to introduce \textquotedblleft
purely virtual\textquotedblright\ particles by tweaking the usual
diagrammatics in a certain way \cite{PVparticles}. The concept leads to a
unitary and renormalizable\ theory of quantum gravity \cite{LWGrav}, whose
main prediction (in the realm of current or planned observations \cite%
{CMBStage4}) is a very constrained window ($4/10000\lesssim r\lesssim 3/1000$%
) for the value of the tensor-to-scalar ratio $r$ \cite{ABP} of primordial
fluctuations. It is delimited from above by the prediction of the
Starobinsky $R+R^{2}$ model, and from below by the properties of the purely
virtual particles themselves\footnote{%
This is also the reason why we have concentrated on the Starobinsky scenario
in discussing inflation. It is apparent that high-energy physics favors that
option over the others.}.

The significance of virtuality in quantum physics suggests that maybe the
entire universe will one day become purely virtual, \textit{de facto}. As
far as we can tell today, the only possibility to make \textquotedblleft
virtuality win over reality\textquotedblright\ is to envision scenarios
where mechanisms like the ones mentioned above make the accelerated
expansion of the universe prevail. Then, at some point, the relatively tiny
on-shell contributions to the particle propagators will become devoid of
practical consequences.

\section{Conclusions}

\label{conclusions}\setcounter{equation}{0}

We have investigated the effects of the expansion of the universe in several
systems, and inquired whether the ultimate fate of the universe is to reach
a state of \textquotedblleft total dilution\textquotedblright , where all
the unstable particles have decayed and the stable ones are so widely
separated from one another that, due to the accelerated expansion, they are
unable to exchange physical signals for the rest of time. Then they remain
virtual forever, and likely entangled, since they cannot interact with
macroscopic objects that can collapse their wave functions. Evocatively, we
could call the final state \textquotedblleft cosmic
virtuality\textquotedblright , or \textquotedblleft eternal
entanglement\textquotedblright .

Homogeneity is not necessary for the expansion to prevail. The
Majumdar-Papapetrou geometries provide nonhomogeneous systems where the
gravitational force is balanced by the electrostatic repulsion. The
Kastor-Traschen extension to a nonvanishing cosmological constant
illustrates the evolution toward cosmic virtuality through an unlimited
expansion. This raises the question whether the expansion prevails any time
the gravitational attraction is balanced by an opposing force. We have shown
that it is not so. In white dwarfs and neutron stars, for example, the
degeneracy pressure stops the gravitational collapse. However, it is not
powerful enough to trigger an unlimited expansion. It would have been in the
early universe, when no celestial bodies existed.

Precisely, once the expansion of the universe is taken into account, the
equilibrium configuration that allows for the presence of stars exists only
if the centrifugal force due to the expansion is smaller than the
gravitational force (at the border of the star, for definiteness). This has
been true since the time of last scattering, but not during inflation.

The question remains: will the universe end in a leopard spot pattern
consisting of isolated regions (e.g., galaxy clusters) that cannot
physically communicate, yet within which macroscopic objects will continue
to exist forever, possibly exhibiting intriguing internal dynamics? Or will
those isles also undergo \textquotedblleft dismemberment\textquotedblright\
from within, akin to the evaporation of black holes? While classical physics
tends to favor the former possibility, quanta are notorious for defying
absolute stability, potentially leading to the latter scenario, which is the
one we favor. In any case, we must defer the final verdict to further
investigations, as a definitive conclusion eludes us at present.

The metrics we have studied can also be used to treat perennial
inhomogeneities, such as primordial black holes amidst the cosmic background
radiation. A hypothesis worth exploring is the idea that the universe might
have one or more \textquotedblleft centers\textquotedblright , meaning,
celestial bodies that have existed before inflation. Studying the impact of
these options on inflation, we have shown that there is room for nontrivial
inhomogeneities without contradicting the established knowledge on
primordial cosmology.

\vskip2 truecm \noindent {\large \textbf{Acknowledgments}}

The author is grateful to U. Aglietti, D. Comelli, E. Gabrielli and M. Piva
for helpful discussions.

\vskip 1.5truecm

\noindent {\textbf{\huge Appendices}} \renewcommand{\thesection}{%
\Alph{section}} \renewcommand{\theequation}{\thesection.\arabic{equation}} %
\setcounter{section}{0}

\section{Higher-order corrections in inflation with primordial black holes}

\label{higher}\setcounter{equation}{0}

In this appendix we give the corrections of order $r_{g}^{2}/u^{2}$ to the
solution (\ref{inflat}) of the equations (\ref{staroeq}) for inflation with
a pre-existing black hole. We find%
\begin{eqnarray*}
\mathrm{d}s^{2} &\text{:}&\text{\qquad }-\frac{r_{g}^{2}}{u^{2}}\left( h_{1}+%
\frac{h_{2}}{m_{\phi }^{2}u^{2}}+\frac{h_{3}}{m_{\phi }^{4}u^{4}}\right) (%
\mathrm{d}t^{2}+a(t)^{2}\mathrm{d}r^{2}+a(t)^{2}r^{2}\mathrm{d}\theta
^{2}+a(t)^{2}r^{2}\sin ^{2}\theta \mathrm{d}\phi ^{2}) \\
&&\qquad +\frac{r_{g}^{2}H^{2}}{2m_{\phi }^{2}u^{2}}\left( 7-3\alpha
^{2}\right) \mathrm{d}t^{2}+\mathcal{O}\left( \frac{r_{g}^{2}}{u^{2}}\frac{1%
}{m_{\phi }^{6}u^{6}}\right) +\mathcal{O}\left( \frac{r_{g}^{3}}{u^{3}}%
\right) , \\
\phi (t,r) &\text{:}&\text{ }\qquad \frac{r_{g}^{2}}{4\sqrt{3\pi G}H\alpha
u^{2}}\left( h_{4}+\frac{3h_{2}H-\dot{h}_{2}}{m_{\phi }^{2}u^{2}}+\frac{%
5h_{3}H-\dot{h}_{3}}{m_{\phi }^{4}u^{4}}\right) +\mathcal{O}\left( \frac{%
r_{g}^{2}}{u^{2}}\frac{1}{m_{\phi }^{6}u^{6}}\right) +\mathcal{O}\left( 
\frac{r_{g}^{3}}{u^{3}}\right) ,
\end{eqnarray*}%
where%
\begin{equation*}
h_{4}=h_{1}H-\dot{h}_{1}-\frac{3H^{2}}{2m^{2}}\left( 2\alpha \dot{\alpha}%
+H(1+5\alpha ^{2}-6\alpha ^{4})\right) ,
\end{equation*}%
and $h_{1,2,3}$ are functions of $\alpha $ that solve certain second order
differential equations. We do not report those equations here, since they
are quite involved. We just point out that they can be solved by expanding
in powers of $\alpha $. The lowest orders of the solutions read%
\begin{eqnarray*}
h_{1} &=&\frac{3}{8}\left( 1-3\alpha +\frac{15}{4}\alpha ^{2}-\frac{19}{6}%
\alpha ^{3}\right) +\mathcal{O}\left( \alpha ^{4}\right) , \\
h_{2} &=&-\frac{\alpha ^{2}}{4}\left( 1+\frac{4}{3}\alpha \right) +\mathcal{O%
}\left( \alpha ^{4}\right) ,\qquad h_{3}=-\frac{2}{5}\alpha ^{2}\left( 1+%
\frac{77}{15}\alpha \right) +\mathcal{O}\left( \alpha ^{4}\right) .
\end{eqnarray*}

\section{Equations of fluid dynamics in general relativity}

\label{fluidGR}\setcounter{equation}{0}

In this appendix we define a fluid and give its equations in general
relativity.

Consider a system of particles labeled by some index $a$. We denote their
positions and velocities by $\mathbf{r}_{a}(t)$ and $\mathbf{v}_{a}(t)=%
\mathrm{d}\mathbf{r}_{a}(t)/\mathrm{d}t$, respectively. When we switch to
the continuum, $\mathbf{v}_{a}(t)$ becomes a field of velocities $\mathbf{v}(%
\mathbf{r},t)$. Precisely, $\mathbf{v}(\mathbf{r},t)$ stands for the
velocity of the fluid element located in $\mathbf{r}$ at time $t$.

It is convenient to define the particle trajectories $x_{a}^{\mu }(t)=(t,%
\mathbf{r}_{a}(t))$ in spacetime. The spacetime velocity $v_{a}^{\mu }(t)=%
\mathrm{d}x_{a}^{\mu }(t)/\mathrm{d}t=(1,\mathbf{v}_{a}(t))$ becomes $v^{\mu
}(x)\equiv (1,\mathbf{v}(t,\mathbf{r}))$ in the continuum limit. The four
velocity $U_{a}^{\mu }(t)=v_{a}^{\mu }(t)/\sigma _{a}(t)$, where $\sigma
_{a}(t)=\sqrt{g_{\rho \sigma }(t,\mathbf{r}_{a})v_{a}^{\rho
}(t)v_{a}^{\sigma }(t)}$, becomes a four vector $U^{\mu }(x)=v^{\mu }/\sqrt{%
g_{\rho \sigma }v^{\rho }v^{\sigma }}$ that satisfies the condition $g_{\mu
\nu }U^{\mu }U^{\nu }=1$.

A fluid is described by the fields of velocities $v^{\mu }$ and $U^{\mu }$,
certain densities $\varepsilon $, $\rho $ of energy and mass (defined below)
and a pressure $P$, such that the current of mass $J^{\mu }$ and the
energy-momentum tensor $T^{\mu \nu }$ read 
\begin{equation}
J^{\mu }=\rho U^{\mu },\qquad T^{\mu \nu }=(\varepsilon +P)U^{\mu }U^{\nu
}-Pg^{\mu \nu }.  \label{tmunu}
\end{equation}%
These definitions show that $\rho =J_{\mu }U^{\mu }$, $\varepsilon =T_{\mu
\nu }U^{\mu }U^{\nu }$ and $P=(\varepsilon -T_{\mu }^{\mu })/3$ are scalars.

Let $\rho _{m}=\mathrm{d}m/\mathrm{d}^{3}\mathbf{r}$ denote the density of
(rest) mass in a given reference frame. The conservation of mass gives the
continuity equation 
\begin{equation}
\partial _{\mu }\left( \rho _{m}v^{\mu }\right) =\frac{\partial \rho _{m}}{%
\partial t}+\bm{\nabla }\cdot \left( \rho _{m}\mathbf{v}\right) =0.
\label{CC}
\end{equation}%
Identifying the scalar $\rho $ with 
\begin{equation}
\rho =\rho _{m}\frac{\sqrt{g_{\mu \nu }v^{\mu }v^{\nu }}}{\sqrt{-g}},
\label{rot}
\end{equation}%
(\ref{CC}) can be written in a manifestly covariant form as the conservation
of the current $J^{\mu }$:%
\begin{equation}
D_{\mu }J^{\mu }=\frac{1}{\sqrt{-g}}\partial _{\mu }\left( \sqrt{-g}\rho
U^{\mu }\right) =\frac{1}{\sqrt{-g}}\partial _{\mu }\left( \rho _{m}v^{\mu
}\right) =0.  \label{C}
\end{equation}

Focusing on an infinitesimal fluid element, it is sometimes convenient to
switch to the \textquotedblleft proper\textquotedblright\ frame, where the
fluid element is at rest. There, $v^{\mu }=(1,\mathbf{0})$, so (\ref{tmunu})
and (\ref{rot}) give $J^{0}=\rho /\sqrt{g_{00}}=\rho _{m}/\sqrt{-g}$. This
means that $\rho $ can be defined as the scalar that coincides with the
density of rest mass $\rho _{m}$ multiplied by $\sqrt{g_{00}}/\sqrt{-g}$ in
the proper frame.

As a consequence of the Einstein equations%
\begin{equation}
R_{\mu \nu }-\frac{1}{2}g_{\mu \nu }R-\Lambda g_{\mu \nu }=8\pi GT_{\mu \nu
},  \label{E}
\end{equation}%
the energy-momentum tensor $T^{\mu \nu }$ is also conserved ($D_{\nu }T^{\mu
\nu }=0$). The conservation of $T^{\mu \nu }$ encodes the equations of
motion of the fluid, as we show below.

What distinguishes a specific fluid from another is the \textquotedblleft
equation of state\textquotedblright\ $P=P(\varepsilon )$, or $\varepsilon
=\varepsilon (P)$, which relates $\varepsilon $ and $P$. Given the equation
of state, the compatibility between the continuity equation (\ref{C}) and
the conservation of $T^{\mu \nu }$ gives a general formula relating $\rho $
to $\varepsilon $. Precisely, using $U_{\mu }D_{\nu }U^{\mu }=0$, which
follows from $g_{\mu \nu }U^{\mu }U^{\nu }=1$, we find%
\begin{equation}
U_{\mu }D_{\nu }T^{\mu \nu }=\frac{\mathrm{d}\varepsilon }{\mathrm{d}\rho }%
D_{\mu }J^{\mu }+\left( \varepsilon +P-\rho \frac{\mathrm{d}\varepsilon }{%
\mathrm{d}\rho }\right) (D_{\mu }U^{\mu }),  \label{cc}
\end{equation}%
wherefrom it is evident that (\ref{C}) and $D_{\nu }T^{\mu \nu }=0$ imply 
\begin{equation}
\varepsilon +P=\rho \frac{\mathrm{d}\varepsilon }{\mathrm{d}\rho }.
\label{important}
\end{equation}%
Assuming the equation of state $P=P(\varepsilon )$, this relation can be
integrated to give%
\begin{equation}
\rho (\varepsilon )=\rho _{0}\exp \left( \int_{\varepsilon
_{0}}^{\varepsilon }\frac{\mathrm{d}\varepsilon ^{\prime }}{\varepsilon
^{\prime }+P(\varepsilon ^{\prime })}\right) ,  \label{rotilde}
\end{equation}%
where $\rho _{0}$ is the value at some reference energy $\varepsilon _{0}$.

It is important to note that (\ref{rotilde}) is not an equation of state,
specific to the fluid, but a universal formula. Ultimately, the fluid is
solely identified by its equation of state $P=P(\varepsilon )$. Formula (\ref%
{rotilde}) can be viewed as the definition of $\varepsilon $ from $\rho $.

Using $U_{\mu }D_{\nu }U^{\mu }=0$ again and $p^{\mu }=mU^{\mu }$, the
identity\ $D_{\nu }T^{\nu \mu }-U^{\mu }U_{\rho }D_{\nu }T^{\nu \rho }=0$
gives%
\begin{equation}
(\varepsilon +P)U^{\nu }D_{\nu }p^{\mu }=m(g^{\mu \nu }-U^{\nu }U^{\mu
})D_{\nu }P.  \label{flM}
\end{equation}%
These are the equations of motion of the fluid, derived from the
conservation of $T^{\mu \nu }$. Note that only three equations are
independent, since contracting with $U_{\mu }$ gives $0=0$.

To obtain a more explicit form of the equations, we multiply (\ref{flM}) by $%
\sqrt{g_{\rho \sigma }v^{\rho }v^{\sigma }}$, divide by $\varepsilon +P$,
lower the index $\mu $, use\footnote{%
Since $\mathbf{p}=\partial L/\partial \mathbf{r}$ and $\mathbf{r}$ has upper
space indices, $\mathbf{r}=(x^{i})$, the vector $\mathbf{p}$ has lower space
indices: $\mathbf{p}=(p_{i})$. The minus sign in front of $\mathbf{p}$ in $%
p_{\mu }=(p_{0},-\mathbf{p})$ is easily checked in flat space.} $p_{\mu
}=mU_{\mu }=(p_{0},-\mathbf{p})$ and finally specialize to a space index $%
\mu =i$. The result is%
\begin{equation}
\frac{\partial \mathbf{p}}{\partial t}+(\mathbf{v}\hspace{0.01in}\cdot %
\bm{\nabla })\mathbf{p}=v^{\mu }\partial _{\mu }\mathbf{p=f}_{G}+\mathbf{f}%
_{P},  \label{fleqmP}
\end{equation}%
where 
\begin{equation}
\mathbf{f}_{G}=-v^{\nu }\Gamma _{\nu \rho }^{\mu }p_{\mu }\bm{\nabla }%
x^{\rho },\qquad \mathbf{f}_{P}=-\frac{m\sqrt{g_{\mu \nu }v^{\mu }v^{\nu }}}{%
\varepsilon +P}\bm{\nabla }P-\frac{\mathbf{p}}{\varepsilon +P}v^{\mu
}\partial _{\mu }P  \label{forces}
\end{equation}%
are the gravitational force and the force due to the pressure, respectively,
while $\bm{\nabla }x^{\rho }$ are just the vectors having components $\delta
_{i}^{\rho }$.

In the next appendix we perform the switch from a set of particles to a
fluid in detail. We show that (\ref{fleqmP}) is the fluid version of the
equation of motion $\mathrm{d}\mathbf{p/}\mathrm{d}t=\mathbf{f}$ obeyed by
each particle individually: the left-hand side is the fluid version of $%
\mathrm{d}\mathbf{p/}\mathrm{d}t$ and the right-hand side is the fluid
version of the total force $\mathbf{f}$.

Gathering the pieces of information learned so far, the (Einstein E,
continuity C and motion M) equations of fluid dynamics in general relativity
are 
\begin{eqnarray}
&&\text{(E):\quad }R_{\mu \nu }-\frac{1}{2}g_{\mu \nu }R-\Lambda g_{\mu \nu
}=8\pi GT_{\mu \nu },\qquad T^{\mu \nu }=(\varepsilon +P)U^{\mu }U^{\nu
}-Pg^{\mu \nu },  \notag \\
&&\text{(C):\quad }D_{\mu }(\rho U^{\mu })=0,\qquad \qquad \text{(M):\quad }%
\varepsilon U^{\nu }D_{\nu }U^{\mu }=D^{\mu }P-U^{\nu }D_{\nu }(PU^{\mu }),
\label{fluiddynGR}
\end{eqnarray}%
(M) just being an alternative way of writing (\ref{flM}).

\section{From a set of particles to a fluid}

\label{switch}\setcounter{equation}{0}

In this appendix we explain how to switch from a set of particles to a
fluid. As before, the particles\ are labeled by the index $a$, with
positions $\mathbf{r}_{a}(t)$, velocities $\mathbf{v}_{a}(t)$ and momenta $%
\mathbf{p}_{a}(t)=\partial L/\partial \mathbf{r}_{a}(t)$. They are subject
to forces $\mathbf{f}_{a}(t)$ and obey the equations of motion 
\begin{equation}
\frac{\mathrm{d}\mathbf{p}_{a}(t)}{\mathrm{d}t}=\mathbf{f}_{a}(t).
\label{dp}
\end{equation}%
We do not need to specify the Lagrangian here, since the arguments apply to
an arbitrary $L$.

When we switch to the continuum, $\mathbf{v}_{a}(t)$ becomes the field of
velocities $\mathbf{v}(\mathbf{r},t)$, and $\mathbf{p}_{a}(t)$ becomes a
field of momenta $\mathbf{p}(\mathbf{r},t)$. The formula relating $\mathbf{p}%
(\mathbf{r},t)$ to $\mathbf{v}(\mathbf{r},t)$ and $\mathbf{r}$ is obtained
by performing the conversion on the formula that holds for single particles, 
$\mathbf{p}_{a}(t)=\partial L/\partial \mathbf{r}_{a}(t)$.

We want to switch from (\ref{dp}) to the equations of motion of the fluid.
Let $\rho _{n}(t,\mathbf{r})=\mathrm{d}n/\mathrm{d}^{3}\mathbf{r}$ denote
the number of particles per unit volume (in a generic reference frame, which
we do not need to specify). Consider a set of particles surrounded by a
closed surface $S$. We denote the interior of $S$ by $V$ and assume that $S$
deforms in time so that no particle exits from $S$ nor enters into it. The
total momentum of the particles contained in $S$ is%
\begin{equation}
\mathbf{p}_{S}=\int_{V}\rho _{n}\mathbf{p}\hspace{0.01in}\mathrm{d}^{3}%
\mathbf{r}.  \label{ps}
\end{equation}%
Its derivative with respect to time is the force acting on $V$. We have%
\begin{equation}
\frac{\mathrm{d}\mathbf{p}_{S}}{\mathrm{d}t}=\int_{V}\left( \frac{\partial
\rho _{n}}{\partial t}\mathbf{p}+\rho _{n}\frac{\partial \mathbf{p}}{%
\partial t}\right) \hspace{0.01in}\mathrm{d}^{3}\mathbf{r}+\int_{S}\hspace{%
0.01in}(\rho _{n}\mathbf{v}\cdot \hat{\mathbf{n}})\mathbf{p}\mathrm{d}\sigma
,  \label{dps0}
\end{equation}%
where $\mathrm{d}\sigma $ is the surface element on $S$. The last integral
arises from the assumption that $S$ encloses the same set of particles as
time passes: the variation of the volume $V$ is locally $\mathrm{d}V=\mathbf{%
v}\cdot \hat{\mathbf{n}\hspace{0.01in}}\mathrm{d}\sigma \mathrm{d}t$ due to
the deformation of the surface $S$ that surrounds it.

Converting the surface integral of (\ref{dps0}) to a volume one by means of
Gauss' theorem, we obtain%
\begin{equation}
\frac{\mathrm{d}\mathbf{p}_{S}}{\mathrm{d}t}=\int_{V}\rho _{n}\left( \frac{%
\partial \mathbf{p}}{\partial t}+(\mathbf{v}\hspace{0.01in}\cdot \bm{\nabla }%
)\mathbf{p}\right) \hspace{0.01in}\mathrm{d}^{3}\mathbf{r}+\int_{V}\left( 
\frac{\partial \rho _{n}}{\partial t}+\bm{\nabla }\cdot (\rho _{n}\mathbf{v}%
)\right) \mathbf{p}\hspace{0.01in}\mathrm{d}^{3}\mathbf{r}.  \label{dps}
\end{equation}%
\ 

En passant, note that the continuity equation (\ref{CC})\ follows from the
same line of reasoning. Consider a fluid composed of particles of the same
mass, $m$. If we set $\mathbf{p}\rightarrow m$ in (\ref{ps}), $\mathbf{p}%
_{S} $ is replaced by the mass $M_{S}$ contained in $V$ (due to $\rho
_{m}=m\rho _{n}$), which is constant. Then the left-hand side of (\ref{dps})
and the first integral on the right-hand side vanish. Being the volume $V$
arbitrary, the remaining integral with $\mathbf{p}\rightarrow m$ gives (\ref%
{CC}).

We can repeat the argument with anything we want in (\ref{ps}), instead of $%
\mathbf{p}$. The outcome is a general rule stating that when we switch from
a set of particles to a fluid, the total derivative with respect to time
undergoes the replacement%
\begin{equation}
\frac{\mathrm{d}}{\mathrm{d}t}\rightarrow \frac{\partial }{\partial t}+%
\mathbf{v}\hspace{0.01in}\cdot \bm{\nabla }=v^{\mu }\partial _{\mu }.
\label{rula}
\end{equation}%
For example, if we apply this rule to $x_{a}^{\mu }(t)\rightarrow x^{\mu }$,
we find $v_{a}^{\mu }(t)=\mathrm{d}x_{a}^{\mu }(t)/\mathrm{d}t\rightarrow
v^{\nu }\partial _{\nu }x^{\mu }=v^{\mu }$, as expected.

Finally, from (\ref{dp}), the force acting on the volume $V$ is 
\begin{equation}
\sum_{a\in V}\mathbf{f}_{a}=\int_{V}\rho _{n}\mathbf{f}\hspace{0.01in}%
\mathrm{d}^{3}\mathbf{r}.  \label{totf}
\end{equation}%
Equating this expression to (\ref{dps}), using the continuity equation, and
recalling that the volume $V$ is arbitrary, we obtain the equation of motion 
\begin{equation}
\frac{\partial \mathbf{p}}{\partial t}+(\mathbf{v}\hspace{0.01in}\cdot %
\bm{\nabla })\mathbf{p}=\mathbf{f},  \label{eomdust}
\end{equation}%
which is just (\ref{dp}) with the replacements $\mathbf{p}_{a}(t)\rightarrow 
\mathbf{p}(t,\mathbf{r})$, $\mathbf{f}_{a}(t)\rightarrow \mathbf{f}(t,%
\mathbf{r})$ and (\ref{rula}). In the cases of interest to us, the total
force $\mathbf{f}$ on the right-hand side is the sum of the forces given in
formula (\ref{forces}) (plus the centrifugal force due to the expansion of
the universe, if we use static coordinates).

\section{Dust}

\label{dust}

In this appendix we study the fluid equations (\ref{eomdust}) and (\ref{flM}%
) in the simpler case of zero pressure (dust). In the next appendix we
switch to the Fermi fluid.

Dust is a set of particles of mass $m$ described by the action%
\begin{equation*}
S=-\int \mathrm{d}t\sum_{a}m\sqrt{g_{\mu \nu }(t,\mathbf{r}%
_{a}(t))v_{a}^{\mu }(t)v_{a}^{\nu }(t)}=-\int \mathrm{d}^{4}x\sum_{a}m\delta
^{(3)}(\mathbf{r}-\mathbf{r}_{a}(t))\sqrt{g_{\mu \nu }(x)v_{a}^{\mu
}(t)v_{a}^{\nu }(t)}.
\end{equation*}

The density of mass is 
\begin{equation*}
\rho _{m}=m\sum_{a}\delta ^{(3)}(\mathbf{r}-\mathbf{r}_{a}(t)),
\end{equation*}%
while the energy-momentum tensor reads 
\begin{equation*}
T^{\mu \nu }(x)=-\frac{2}{\sqrt{-g(x)}}\frac{\delta S}{\delta g_{\mu \nu }(x)%
}=\sum_{a}\frac{m\delta ^{(3)}(\mathbf{r}-\mathbf{r}_{a}(t))v_{a}^{\mu
}(t)v_{a}^{\nu }(t)}{\sqrt{-g(x)}\sqrt{g_{\rho \sigma }(x)v_{a}^{\rho
}v_{a}^{\sigma }}}.
\end{equation*}

When we switch to the fluid ($v_{a}^{\mu }(t)\rightarrow v^{\mu }(x)$), we
find 
\begin{equation*}
T^{\mu \nu }\rightarrow \varepsilon U^{\mu }U^{\nu },\qquad \varepsilon
=\rho _{m}\frac{\sqrt{g_{\mu \nu }v^{\mu }v^{\nu }}}{\sqrt{-g}}=\rho ,\qquad
P=0,
\end{equation*}%
having used (\ref{rot}). The identity (\ref{important}) is trivially
satisfied, since $\varepsilon =\rho $, $P=0$.

The $a$-th particle moves according to the geodesic equation%
\begin{equation}
\frac{\mathrm{d}U_{a}^{\mu }(t)}{\mathrm{d}t}+\Gamma _{\nu \rho }^{\mu }(t,%
\mathbf{r}_{a}(t))v_{a}^{\nu }(t)U_{a}^{\rho }(t)=0,  \label{geoa}
\end{equation}%
from which we can read the force $\mathbf{f}_{a}$ of (\ref{dp}). Repeating
the procedure applied for deriving (\ref{fleqmP}), i.e., using $p_{\mu
}=(p_{0},-\mathbf{p})=g_{\mu \nu }p^{\mu }$, which implies $p_{i}=-g_{i\mu
}p^{\mu }$, and specializing to a space index $\mu =i$, we find 
\begin{equation*}
\frac{\mathrm{d}\mathbf{p}_{a}(t)}{\mathrm{d}t}=-\Gamma _{\rho \nu }^{\mu
}(t,\mathbf{r}_{a}(t))p_{a\mu }(t)v_{a}^{\nu }(t)\bm{\nabla }x^{\rho }=%
\mathbf{f}_{a}(t).
\end{equation*}%
Switching from a set of particles to a fluid by means of (\ref{rula}), we
obtain (\ref{eomdust}) with $\mathbf{f}=\mathbf{f}_{G}$, the gravitational
force given in (\ref{forces}). Dividing (\ref{eomdust}) by $\sqrt{g_{\rho
\sigma }v^{\rho }v^{\sigma }}$, we find the spatial components of $U^{\nu
}D_{\nu }p_{\mu }=0$, which is (\ref{flM}) at $P=0$.

In particular, the equations of motion of dust, $U^{\nu }D_{\nu }U^{\mu }=0$%
, are just the geodesic equations (\ref{geoa}) switched to the continuum.

\section{Ideal Fermi fluid at zero temperature}

\label{degeneracy}\setcounter{equation}{0}

In this appendix we consider a fluid of fermions at zero temperature,
assuming that the interactions among the particles are negligible. We derive
the degeneracy pressure $P$ and the equation of state. We also show how to
switch from the equations of a system of particles to the ones of a fluid in
a certain limit, where it is possible to proceed in a direct way. For more
details, see \cite{Stars,Landau5}.

\subsection{Degeneracy pressure and equation of state}

A Fermi fluid at zero temperature consists of particles with nonzero
energies, filling the Fermi surface. By interpreting those energies as
internal, and viewing a fluid element as a whole,\ we can determine whether
it is at rest or in motion, and define the field of velocities $\mathbf{v}(t,%
\mathbf{r})$, as well as $\varepsilon $, $\rho $ and the pressure $P$.

Referring to the formulas (\ref{tmunu}), we recall that $\rho $, $%
\varepsilon $ and $P$ are scalars, so it is enough to work out their
relations in flat space. Moreover, we can focus on a particular fluid
element and switch to its proper\ frame, where $v^{\mu }=U^{\mu }=(1,\mathbf{%
0})$. Then $\rho $ coincides with the density of mass $\rho _{m}$, by
formula (\ref{rot}). We can read $\varepsilon $ and $P$ from the
energy-momentum tensor $T^{\mu \nu }=$ diag$(\varepsilon ,P,P,P)$.
Specifically, $\varepsilon $ is the energy density.

The interval is $\mathrm{d}s^{2}=\mathrm{d}t^{2}-\mathrm{d}\mathbf{r}^{2}$,
the space momentum of the fluid element is $\mathbf{p}=\partial L/\partial 
\mathbf{r}=(p_{i})$ and the commutation relations are $[\hat{p}_{i},\hat{x}%
^{j}]=-i\delta _{i}^{j}$. The number of quantum states per unit phase-space
volume is 
\begin{equation*}
\mathrm{d}n=(2s+1)\frac{\mathrm{d}^{3}\mathbf{p}\hspace{0.01in}\mathrm{d}^{3}%
\mathbf{r}}{h^{3}}\theta (p_{F}-p),
\end{equation*}%
where $s$ is the particle spin and $p_{F}$ is the Fermi momentum. For
neutrons, protons and electrons, the phase space density $\rho _{\text{ph}%
}(p)$ and the density $\rho $ are 
\begin{equation}
\rho _{\text{ph}}(p)=\frac{\mathrm{d}n}{\mathrm{d}^{3}\mathbf{p}\hspace{%
0.01in}\mathrm{d}^{3}\mathbf{r}}=\frac{2}{h^{3}}\theta (p_{F}-p),\qquad \rho
=\rho _{m}=\frac{\mathrm{d}m}{\hspace{0.01in}\mathrm{d}^{3}\mathbf{r}}=m\int
\rho _{\text{ph}}(p)\mathrm{d}^{3}\mathbf{p}=\frac{8\pi mp_{F}^{3}}{3h^{3}},
\label{density}
\end{equation}%
respectively.

The energy density $\varepsilon $ is obtained by summing the energies $\sqrt{%
m^{2}+\mathbf{p}^{2}}$ of each particle, which gives 
\begin{equation}
\varepsilon =\int \sqrt{m^{2}+\mathbf{p}^{2}}\rho _{\text{ph}}(p)\mathrm{d}%
^{3}\mathbf{p}=\frac{8\pi }{h^{3}}\int_{0}^{p_{F}}p^{2}\mathrm{d}p\sqrt{%
m^{2}+p^{2}}.  \label{eps}
\end{equation}

To calculate the degeneracy pressure $P$, we proceed as follows. A particle
of momentum $\mathbf{p}$ colliding on an infinitesimal surface $\mathrm{d}S$
transfers a momentum $\Delta p=-2\mathbf{p}\cdot \hat{\mathbf{n}}$ to it,
where $\hat{\mathbf{n}}$ is the normal to the surface. The number of
particles of momentum $\mathbf{p}$ colliding per unit time is $\mathrm{d}%
n_{c}/\mathrm{d}t=-\rho _{\text{ph}}(p)\mathrm{d}^{3}\mathbf{p}(\mathbf{v}%
\cdot \hat{\mathbf{n}})\mathrm{d}S$. Integrating on the hemisphere H of
incident particles, we obtain 
\begin{equation}
P=\int_{\text{H}}\Delta p\frac{\mathrm{d}n_{c}}{\mathrm{d}S\mathrm{d}t}%
=2\int_{\text{H}}(\mathbf{p}\cdot \hat{\mathbf{n}})(\mathbf{v}\cdot \hat{%
\mathbf{n}})\rho _{\text{ph}}(p)\mathrm{d}^{3}\mathbf{p}=\frac{8\pi }{3h^{3}}%
\int_{0}^{p_{F}}\frac{p^{4}\mathrm{d}p}{\sqrt{m^{2}+p^{2}}}.  \label{pdega}
\end{equation}%
Finally, we find the relations%
\begin{equation}
\varepsilon +P=\frac{8\pi p_{F}^{3}}{3h^{3}}\varepsilon _{F},\qquad P=\frac{%
\pi }{3h^{3}}\left[ \varepsilon _{F}p_{F}(2\varepsilon
_{F}^{2}-5m^{2})+3m^{4}\ln \frac{\varepsilon _{F}+p_{F}}{m}\right] ,\qquad
\varepsilon _{F}=\sqrt{m^{2}+p_{F}^{2}},  \label{state}
\end{equation}%
which give the equation of state $P=P(\varepsilon )$ implicitly. Similarly,
using (\ref{density}) to eliminate $p_{F}$, we can also find $\varepsilon $
as a function of $\rho $. We stress again that these relations are valid in
an arbitrary frame and with an arbitrary metric, since $\rho $, $P$ and $%
\varepsilon $ are scalars.

The integral formulas (\ref{eps}) and (\ref{pdega}) show that $\varepsilon $
and $P$ tend to zero if and only if $p_{F}$ tends to zero. Moreover, the
ratio of the integrals shows that $P(\varepsilon )/\varepsilon $ tends to
zero when $\varepsilon $ tends to zero, which was assumed in section \ref%
{neutron}.

Now we check the identity (\ref{important}). Comparing the first formula of (%
\ref{state}) with the second of (\ref{density}), we obtain $\varepsilon
+P=\rho \varepsilon _{F}/m$. Moreover, (\ref{eps}) and (\ref{density}) give $%
m\mathrm{d}\varepsilon /\mathrm{d}p_{F}=\varepsilon _{F}\mathrm{d}\rho /%
\mathrm{d}p_{F}$, which implies $\mathrm{d}\varepsilon /\mathrm{d}\rho
=\varepsilon _{F}/m$. Combining these results, we find $\varepsilon (\rho
)+P(\varepsilon (\rho ))=\rho \mathrm{d}\varepsilon (\rho )/\mathrm{d}\rho $%
, which is (\ref{important}).

A useful formula for the arguments of section \ref{neutron} is obtained by
combining (\ref{density}), (\ref{eps}) and (\ref{pdega}) to work out the
primitive%
\begin{equation}
\int \frac{\mathrm{d}\rho }{\rho }\frac{\mathrm{d}P}{\mathrm{d}\rho }\left( 
\frac{\mathrm{d}\varepsilon }{\mathrm{d}\rho }\right) ^{-1}=\int \frac{%
\mathrm{d}p_{F}}{\rho }\frac{\mathrm{d}P}{\mathrm{d}p_{F}}\frac{\mathrm{d}%
\rho }{\mathrm{d}p_{F}}\left( \frac{\mathrm{d}\varepsilon }{\mathrm{d}p_{F}}%
\right) ^{-1}=\ln \frac{\varepsilon _{F}}{m}=\ln \sqrt{1+\left( \frac{%
3h^{3}\rho }{8\pi m^{4}}\right) ^{2/3}},  \label{primitive}
\end{equation}%
up to an arbitrary additive constant.

In the nonrelativistic limit ($p_{F}\ll m$) we obtain%
\begin{equation}
\varepsilon \simeq \frac{8\pi mp_{F}^{3}}{3h^{3}}=\rho ,\quad P\simeq \frac{%
8\pi p_{F}^{5}}{15mh^{3}}\simeq \frac{\varepsilon ^{5/3}}{20}\left( \frac{%
3h^{3}}{\pi m^{4}}\right) ^{2/3}\ll \varepsilon , \quad \int \frac{\mathrm{d}%
\rho }{\rho } \frac{\mathrm{d}P}{\mathrm{d}\varepsilon } \simeq \frac{1}{2}%
\left( \frac{3h^{3}\rho }{8\pi m^{4}}\right) ^{2/3},  \label{nrstate}
\end{equation}%
and the identity (\ref{important}) is a consequence of $\varepsilon \simeq
\rho $, $P\ll \varepsilon $.

\subsection{Fluid equations in the nonrelativistic Newtonian limit}

We consider the fluid equations (\ref{fleqmP}) in the nonrelativistic limit,
where $P\ll \varepsilon \simeq \rho $, and in the Newtonian approximation (%
\ref{Newt}), where $\mathbf{f}_{G}$ is just Newton's gravitational force.
Using (\ref{rot}), we find, to the lowest nontrivial order in $G$, $P$ and
the velocity, 
\begin{equation}
\mathbf{f}_{P}\simeq -\frac{m\bm{\nabla }P}{\rho _{m}}.  \label{fGnr}
\end{equation}

It is straightforward to derive this expression geometrically. Consider
again a set of particles surrounded by a closed surface $S$, $V$ denoting
its interior. The force exerted by the degeneracy pressure on $V$ is 
\begin{equation}
-\int_{S}P\hspace{0.01in}\hat{\mathbf{n}}\hspace{0.01in}\mathrm{d}\sigma
=-\int_{V}\bm{\nabla }P\hspace{0.01in}\mathrm{d}^{3}\mathbf{r}.  \label{deg}
\end{equation}%
Equating (\ref{dps}) to (\ref{totf}) with $\mathbf{f}\hspace{0.01in}%
\rightarrow \mathbf{f}\hspace{0.01in}_{G}$ plus (\ref{deg}), using the
continuity equation and recalling that the volume $V$ is arbitrary, we obtain%
\begin{equation}
\frac{\partial \mathbf{p}}{\partial t}+(\mathbf{v}\hspace{0.01in}\cdot %
\bm{\nabla })\mathbf{p}\hspace{0.01in}\simeq \hspace{0.01in}\mathbf{f}_{G}-%
\frac{m\bm{\nabla }P}{\rho _{m}},  \label{eq1}
\end{equation}%
in agreement with (\ref{fGnr}). The total force is the sum of two forces $%
\mathbf{f}_{G}$ and $\mathbf{f}_{P}$ that do not talk to each other. Said
differently, the superposition principle holds.

Away from the limit we have just considered, the superposition principle
does not hold, which is why in general relativity we have the more involved
expressions (\ref{fleqmP}) and (\ref{forces}), derived from the covariant
equation (\ref{flM}).

\end{document}